\documentclass{ws-ijmpa}
\usepackage[super,compress]{cite}
\usepackage{graphicx}
\usepackage[T1]{fontenc} 
\usepackage{diagbox}
\usepackage{adjustbox}
\usepackage{subfigure}
\usepackage{multirow}
\usepackage{placeins}

\begin{document}
\markboth{S. Dahbi, J. Choma, B. Mellado, G. Mokgatitswane, X. Ruan, B. Lieberman and T. Celik }{Machine learning approach for the search of resonances with topological features at the LHC}

%
\catchline{}{}{}{}{}
%

\title{Machine learning approach for the search of resonances with topological features at the Large Hadron Collider
}

\author{Salah-Eddine Dahbi$^{*}$, Joshua Choma, Gaogalalwe Mokgatitswane, Xifeng Ruan, \\ Benjamin Lieberman}

\address{School of Physics and Institute for Collider Particle Physics, University of the Witwatersrand,
Johannesburg, Wits 2050, South Africa.\\
salah\_eddine.dahbi@cern.ch$^{*}$}

\author{Bruce Mellado}

\address{School of Physics and Institute for Collider Particle Physics, University of the Witwatersrand,
Johannesburg, Wits 2050, South Africa.\\
iThemba LABS, National Research Foundation, PO Box 722,\\ Somerset West 7129, South Africa.}

\author{Turgay Celik}

\address{School of Electrical and Information Engineering, and Wits Institute of Data Science Physics, University of the Witwatersrand,
Johannesburg, Wits 2050, South Africa.}

\maketitle


\begin{abstract}
The observation of resonances is unequivocal evidence of new physics beyond the Standard Model at the Large Hadron Collider (LHC). So far, inclusive and model dependent searches have not provided evidence of new resonances, indicating these could be driven by subtle topologies. Here, we use machine learning techniques based on weak supervision to perform searches. Weak supervision based on mixed samples can be used to search for resonances with little or no prior knowledge of the production mechanism. Also, it offers the advantage that sidebands or control regions can be used to effectively model backgrounds with minimal reliance on simulations. However, weak supervision alone is found to be highly inefficient in identifying corners of the multi-dimensional space of interest.  Instead, we propose an approach to search for new resonances that involves a classification procedure that is signature and topology based. A combination of weak supervision with Deep Neural Network algorithms are applied following this classification. The performance of this strategy is evaluated on the production of SM Higgs boson decaying to a pair of photons inclusively and in exclusive regions of phase space tailored for specific production modes at the LHC. After verifying the ability of the methodology to extract different SM Higgs boson signal mechanisms, a search for new phenomena in high-mass final states is setup for the LHC.

\end{abstract}

\mbox{}
\clearpage
\section{Introduction}
\label{sec:intro}

After the discovery of a Higgs boson ($h$)\cite{Higgs:1964ia,Englert:1964et,Higgs:1964pj} at the Large Hadron Collider (LHC) by the ATLAS\cite{Aad:2012tfa} and CMS\cite{Chatrchyan:2012ufa} experiments, the Standard Model (SM) of particle physics has exhausted fundamental predictions. The SM is not able to explain a number of phenomena that display overwhelming experimental evidence, such as Dark Matter, the origin of neutrino mass, the matter-anti-matter asymmetry, in addition to a number of theoretical problems. As a result, many theories have been developed to extend the SM and propose a theoretical explanation for phenomena Beyond the Standard Model (BSM). 

In this context, many ongoing experiments, such as those at the LHC, are devoted to exploring physics beyond SM. With the enormous growth in the size of data sets and the complexity of theoretical models in high energy physics (HEP), advanced techniques to detect anomalies are necessitated. These are required to skim through vast experimental data sets and should have the ability to extract small signals in unspecified corners of the phase-space. In order to address this challenge, particle physicists make use of machine learning (ML). 

Searches for new resonances is a pivotal element to the program to unveil new physics beyond the SM. The absence of signals in inclusive searches or benchmark analyses indicates that, should new resonances be hidden in the data, signals are bound to be subtle. Subtlety can be driven by non-trivial couplings of the resonances to the SM particles, leading to topologies that are not adequately covered by benchmark analyses\cite{Englert:2012xt,Djouadi:2013yb}. It can also be driven by the appearance of additional production mechanisms. Therefore, resonances can be missed in corners of the phase-space that are either unexplored or poorly covered by the search. 

Machine Learning can play a significant role in alleviating this problem. Full supervision has proven highly effective in exploring corners of the phase-space of interest, such as those where the SM Higgs boson was observed. However, the effectiveness of this approach is seriously hampered if new physics populates the phase-space differently from the model used as a baseline. The use of weak supervision on the basis of sideband analyses provides the advantage that it can in principle detect topological differences between the potential signal region and the sidebands. However, given the complexity of the phase-space explored and the way backgrounds and potential signals may populate it, weak supervision alone appears to be inefficient. This effect is quantified here using the SM Higgs boson. 

A number of anomalies have been identified based on discrepancies in the production of leptons at the ATLAS and CMS experiments of the LHC\cite{vonBuddenbrock:2019ajh,vonBuddenbrock:2015ema,vonBuddenbrock:2016rmr,Fang:2017tmh,vonBuddenbrock:2017gvy,vonBuddenbrock:2018xar,Sabatta:2019nfg,Hernandez:2019geu,vonBuddenbrock:2020ter,Fischer:2021sqw,Crivellin:2021ubm}. These include a number of final states: opposite-sign, same sign di-leptons, three leptons in the presence and absence of $b$-quarks. These final states appear in corners of the phase-space where different SM processes dominate. 
 
Phenomenological studies indicate that these anomalies can be accommodated by an ansatz composed of two additional scalar fields, $H$ and $S$, where the mass of $H$ is around 270\,GeV and $S$ is SM Higgs-like and has a mass around 150\,GeV\cite{vonBuddenbrock:2016rmr,vonBuddenbrock:2017gvy,vonBuddenbrock:2019ajh,Crivellin:2021ubm}. This serves as a substantiation to search for Higgs-like resonances with a mass around the EW scale, where SM backgrounds are large. Here, topological requirements play an important  role in suppressing backgrounds.

In this paper we extend the applications of ML in HEP to the search for new physics at the LHC, using an approach based on weak supervision learning with topological requirements. The approach is dedicated to extract the potential signals from the experimental data with little prior assumption of the characteristics of the signal. As a proof of principle, the method is tested on the extraction of the SM Higgs boson. The method is intended to identify different phase-space of the resonances, since they are expected to be generated with different production mechanisms at the LHC. This method is adequate to extract more subtle signals in the data with some human intervention.

In this context, we have organised this paper as follows: Section~\ref{sec:MLA} briefly introduces ML classification techniques, namely, supervised learning, unlabelled supervised learning and weak supervision. Their performance will be evaluated in Section~\ref{sec:MLPerform} using the production of the SM Higgs boson decaying to a pair of photons at the LHC. Section~\ref{sec:WSLTopo} proposes an approach for the search for new physics, where weak supervision learning is restricted to the topological configurations of the phase space, such as number of $b$-tagged jets, number of leptons, vector boson fusion topology, etc... Finally, Section~\ref{sec:Concl} summarises the study with suggestions and prospects for the future resonance searches at the LHC.

\section{Machine learning}
\label{sec:MLA}
Machine learning is a field under computer science that is comprised of algorithms that are able to extract insight from data. Most of these algorithms have been successfully used for regression and classification tasks in various fields such as HEP\cite{cohen2018machine}, general industry, science and engineering. Application of these techniques in HEP, which started in the 1990s\cite{albertsson2018machine}, contributed to the discovery of a Higgs boson at the LHC. The use of ML here was driven by full supervision, where the production mechanisms, the corresponding topologies and signatures were well defined by the theory, where the mass of the SM Higgs boson was the only free parameter in the search. By contrast, the search for new resonances, even new Higgs-like bosons, may entail subtle topologies. The use of full supervision here becomes cumbersome, and therefore ineffectual. Such exploration could be accomplished with the assistance of a Lagrangian based on effective interactions, where a considerable number of parameters would be required. 

Here we will be evaluating different approaches as we move away from full supervision towards a full-blown weak supervision approach. In this section, the basic tenets of tools and methodologies that will be implemented in Section~\ref{sec:MLPerform} are summarised.

\subsection{Deep learning}\label{sec:DeepLearning}
The connection structure of neural networks can be classified into deep and shallow, depending on the number of hidden layers implemented. These hidden layers are responsible for transforming the outputs of the preceding $(k-1)^{th}$ layer as input to the succeeding ($k^{th}$) layer, as shown in the following expression:
\begin{equation}
h^{(k)} = f^{(k)}(b^{(k)} + W^{(k)}h^{(k-1)}),~~~     \vec{h}^{(0)}=\vec{x}, 
\label{h}
\end{equation}
where $W^{(k)}$ is the matrix of the weights, $b^{(k)}$ is the bias term, $h^{(k-1)}$ is the output of the $(k-1)^{th}$ hidden  layer, $f^{(k)}$ is the activation function for the $k^{th}$ layer and $\vec{x}$ is the input vector\cite{learning}. The shallow neural networks seem to struggle with discovering useful functions from high dimensional data sets\cite{guest2018deep} as well as learning complex nonlinear functions \cite{baldi2014searching}. Ref.~\citen{albertsson2018machine} explains deep learning as a revolution of neural networks after the improvement of training algorithms and computational power. Deep neural networks (DNNs) display a good performance in pattern recognition tasks\cite{DNN}. Their architectures have proved to be more effective and capable of learning more complex models than shallow neural networks\cite{monica,christian}. The success is also shown in HEP based on Ref.~\citen{baldienhanced}, which indicates that DNNs performed better than shallow ANN and they are good tools for learning discriminatory information in the physics variables\cite{baldi2014searching}. This has been reported in Ref.~\citen{baldi2014searching}, where deep learning methods have demonstrated an improvement of classification metric between signal and background classes by up to $8\%$ above the best current methods. Ref.~\citen{carminati2017calorimetry} shows that deep learning techniques perform better than the traditional techniques in tasks such as particle identification, energy measurement and detector simulation.  Their architectural structure enable them to even learn complex correlations between features with increased depth size of the network\cite{abdughani2019supervised}.
\subsection{Supervised learning}
\label{sec:FSL}
The main purpose of supervised learning is to train a model on fully labelled data where each example $\vec{x_{i}}$ comes with a label $y_{i} \in \{0,1\}$, in a case of binary classification task. This is done with a goal of learning a mapping from $x$ to $y$. The supervised learning model is trained to minimise the loss function which can be in a form of binary cross-entropy:
\begin{equation}
    \ell (y,\hat{y}) = -y\cdot \log \hat{y} + (1-y)\cdot \log (1-\hat{y}), \label{eqn:lossEntropy}
\end{equation}

where $\hat{y}$ is the model output and $y$ is the target output (1 for signal and 0 for background).
The training of this model is done with an objective of finding a function that will be able to estimate the labels or correct categories of a new data set that is presented to the model without labels, \textit{i.e.} testing data to evaluate how well the model generalises. This is achieved by minimizing the loss function to find an optimal solution\cite{dery2017weakly}:
\begin{equation}
    f_{full} = argmin_{f:\mathbb{R}^{n}\rightarrow [0,1]}\sum\limits_{i=1}^N \ell(y_{i}, \hat{y_{i}}),
    \label{eqn:FullSupervision}
\end{equation}
where $f$ is the predictor function, $\hat{y_{i}}$ is the $i^{th}$ model output, $y_{i}$ is the corresponding target output and $\ell$ is the loss for a single example\cite{cheongcs}.
This approach enables the model to produce its own output and continually adjusts its internal states, weights using the back propagation algorithm which follows the principle of gradient descent until the model output approximates the expected output. It is for this reason why supervised learning is called "learning with a teacher"\cite{fyfe2000artificial}. This learning paradigm has shown to be successful when applied to event selection in HEP, when implemented on algorithms such as neural networks\cite{abazov2001search,acosta2005measurement} and SVMs\cite{whiteson2003support} over heuristic selectors\cite{whiteson2009machine}. 

\subsection{Supervised learning with unlabelled data}
\label{sec:UFSL}
 Several studies have focused on the use of unlabelled data to improve the performance of supervised learning algorithms\cite{goldman2000enhancing, chen2019energy,baluja1999probabilistic,nigam2001using,towell1996using,shahshahani1994effect}. This takes place when the training set is composed of $l$ labelled samples $\{(X_{1},\theta_{1}), ..., (X_{l},\theta_{l})\}$ and $u$ unlabelled samples $\{X^{'}_{1},...,X^{'}_{u}\}$.
 Ref.~\citen{baluja1999probabilistic} uses the combination of a statistical technique Expectation Maximization (EM) algorithm and a naive Bayes classifier by learning from both the labelled and unlabelled data. The classifier is trained on the labelled dataset, which is further used to assign labels to the unlabelled sets. Ref.~\citen{chen2019energy} uses a deep learning technique to compensate the new labels for the unlabelled data. This approach is described in  Ref.~\citen{towell1996using} as supervised learning using labelled and unlabelled (SULU) examples and it is used when there are no enough labels. The SULU approach is achieved by finding the centroid of the two sets, $l$ and $u$ in the neighbourhood of the labelled examples. In contrast to this, we combined different signal production mechanisms and assigned them the same class label. The technique takes into account that the mechanisms are completely different. The purpose of this study was to see if the machine learning algorithm is able to detect the individual production mechanisms during the testing phase. 

\subsection{Weakly supervised learning}
\label{sec:WySL}
Training deep learning models requires a large amount of data and can be costly in terms of computational power and having human experts label the training data. Authors in \cite{han2014object} emphasise that the manual labelling of the data is sometimes unreliable. This is where weak supervision\cite{dery2017weakly}, which has sparked interest from many researchers in HEP, comes in. The learning phase during this paradigm takes place on partially/weakly labelled dataset\cite{patrini2016loss}. This is regarded as a cheap form of supervision\cite{han2014object, zhang2016weakly} since it enables the model to learn from a training data that has imprecise labels \cite{dehghani2017neural} and it is mostly applied in computer vision for action recognition \cite{hartmann2012weakly} and object detection \cite{siva2011weakly}.

According to Ref.~\citen{zamani2018theory}, an example of weakly supervised learning would be a case where $\widehat{y} = M(x)$ and $M$ is weak supervision signal, which is used to generate labels for any given input. 
\newpage
A similar approach is implemented in Ref.~\citen{huang2016connectionist}. As a result, this gives a weakly labelled training data 
$ \widehat{\tau} = {(x_{1},\widehat{y}_{1}),(x_{2},\widehat{y}_{2}),\cdot\cdot\cdot,(x_{m},\widehat{y}_{m})}$, where:
\begin{equation}
\widehat{y}_{i} = \left[ M(x_{i1}),M(x_{i2}),\cdot\cdot\cdot,M(x_{in})\right].
\end{equation}

This weakly labelled data is then used by the classifier for the training. Additional benefit of using weak supervision is the decrease in time spent labelling the data manually, which according to Ref.~\citen{yao2016semantic} took 6.8\% of the time spent labelling the fully supervised methods. On the contrary to how the full supervision function is optimised (see equation \ref{eqn:FullSupervision}), weak supervision uses the following equation:
\begin{equation}
    f_{weak} = argmin_{f:\mathbb{R}^{n}\rightarrow [0,1]}\sum\limits_{K} \ell \left( \frac{1}{|K|}\sum\limits_{i\in K}\hat{y_{i}}, y_{K} \right),
    \label{eqn:WeakSupervision}
\end{equation}
where $K$ denotes training data batches and $y_{K}$ is the signal ratio in each batch \cite{cheongcs}.
The performance of weak supervision was compared to that of full supervision in \cite{cohen2018machine,dery2017weakly,siva2011weakly,kuehne2017weakly,huang2016connectionist,yao2016semantic} and weak supervision achieved reasonable results, despite knowing only a fraction of the labels.

\section{ML approaches for the classification of the SM Higgs boson at the LHC }
\label{sec:MLPerform}
The tenets of ML, relevant to this investigation, are discussed in Section~\ref{sec:MLA}. We gradually move from full supervision to weak supervision learning. Compared to full supervision, which requires knowledge of the signal, weak supervision does not require any priory knowledge about the topological features of the signal. The performance of the different ML approaches is evaluated on the SM Higgs boson and background processes with the identical decay products at the LHC, as a benchmark.

\subsection{Datasets and event selection}
\label{sec:MCSim}
The performance of the ML classifiers, described in Section ~\ref{sec:MLA}, are evaluated using simulated Higgs to di-photon ($pp\rightarrow h\rightarrow \gamma\gamma$) events. We focus on the proton-proton collisions at the LHC with a centre-of-mass energy of 13~TeV. 

Background and signal samples were produced using \texttt{MadGraph5$\_$aMC@NLO 2.6.7} with Next-to-Leading Order accuracy in QCD\cite{Alwall:2014hca}. The parton showering and hadronization were simulated with \texttt{PYTHIA 8.2}\cite{Sjostrand:2014zea} using ATLAS A14 event tuning and NNPDF2.3 LO parton distribution function set\cite{Ball:2012cx}. Events were processed with \texttt{Delphes~3}\cite{deFavereau:2013fsa}, which provides an approximate fast simulation of the current ATLAS experiment. Events of interest were simulated by imposing a set of generator-level cuts, where the transverse momentum of the photons is required to be greater than 25\,GeV and the di-photon invariant mass to be between 105 and 160\,GeV. Hadronic jets were reconstructed using the anti-$kt$ algorithm\cite{Cacciari:2008gp} with the radius parameter, $R=0.4$, as implemented in the \texttt{FastJet} 3.2.2\cite{Cacciari:2011ma} package. Jets with $p_{\textrm{T}} > 30$\,GeV and $|\eta| < 4.7$ are considered. In addition, jets originating from bottom quarks are identified as $b$-jets with $b$-tagging algorithms\cite{Vacavant:2009zza}. Reconstructed jets overlapping with photons, electrons or muons in a cone of size $R=0.4$ are removed. Electrons and muons are required to have $p_{\textrm{T}} > 25$\,GeV and $|\eta| < 2.5$. Finally, an inclusive event selection of at least two photons, that have $p_{\textrm{T}} > 25$\,GeV and $|\eta| < 2.37$, is applied.  

\begin{figure}[hbt]
\centering
  \includegraphics[width=0.60\linewidth]{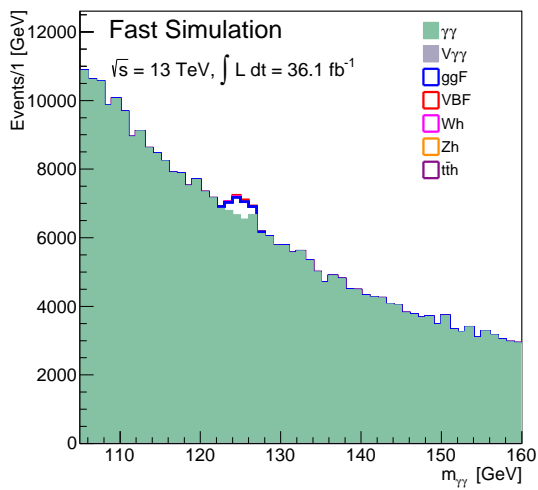}
\caption{Di-photon invariant mass spectrum at 13~TeV of proton-proton collision, corresponds to an integrated luminosity of 36.1\,fb$^{-1}$ of integrated luminosity. The different production mode of the Standard Model Higgs boson samples, are normalised to the cross sections times the di-photon decay branching ratio. }\label{fig1:InvMass}
\end{figure}
\FloatBarrier
\begin{table}[t]
\tbl{The number of entries for each process normalised to the expected events yields.}
{\begin{tabular}{l|ccc}
\hline
\hline
Process & Entries & Event Yields \\
\hline
\hline
$\gamma\gamma$ & 96447
 & 323436\\
V$\gamma\gamma$ & 26613
 & 467\\
ggF & 224835 & 1605\\
VBF & 240085 & 129\\
$Wh$ & 281209 & 41\\
$Zh$ & 169312 & 23\\
$t\bar{t}$h & 160265 & 18\\
  \hline
\hline
\end{tabular}\label{table3:ExpectedYields}}
\end{table}
The dominant backgrounds considered for this study are the non-resonant $\gamma\gamma$, which represent more than 80$\%$ of the total background contribution, and the $V\gamma\gamma$ ($V$=$W,Z$) processes. The different production mechanisms of the SM Higgs boson are treated as separate signal samples.  The production mechanics are the gluon gluon fusion (ggF), vector boson fusion (VBF), and associated productions with a vector boson ($Vh$) or a top–antitop quark pair ($t\bar{t}h$). Samples from each process are normalised according to the prescriptions of the Higgs cross-section working group\cite{deFlorian:2016spz}. Table \ref{table3:ExpectedYields} presents the number of entries of each process normalised to the expected events yields, on which we used 50\% of the events randomly for training the classifier and the remaining 50\% to test the performance of the proposed neural network topology. The $h\rightarrow \gamma\gamma$ branching ratio ($\approx 0.23\%$) is obtained form Ref.~\citen{Denner:2011mq}. The reconstructed di-photon invariant mass spectra, $m_{\gamma\gamma}$, for both the signals and backgrounds are shown in Fig.~\ref{fig1:InvMass}.
\newpage
\subsection{DNN architecture and training strategy}
\label{sec:DNNTMVA}
Events from the simulated dataset are classified using a DNN algorithm, as implemented in the package ROOT\cite{Brun:1997pa} and the TMVA\cite{Therhaag:2009dp} data processing framework. The DNN architecture implemented in this study consists of an input layer, three hidden layers and an outer layer. This architecture is described in Section~\ref{sec:DeepLearning}. The input layer consists of 13 neurons, which represent the kinematic features of the dataset, as described in ~\ref{sec:inputVar}. The three hidden layers are made up of 26, 26 and 13 neurons, respectively. The output layer consists of a single neuron. Two activation functions are used here: a hyperbolic tangent for the hidden layers, and a sigmoid function for the output layer. 

\begin{figure}[hbt!]
\centering
 \includegraphics[width=1.0\linewidth]{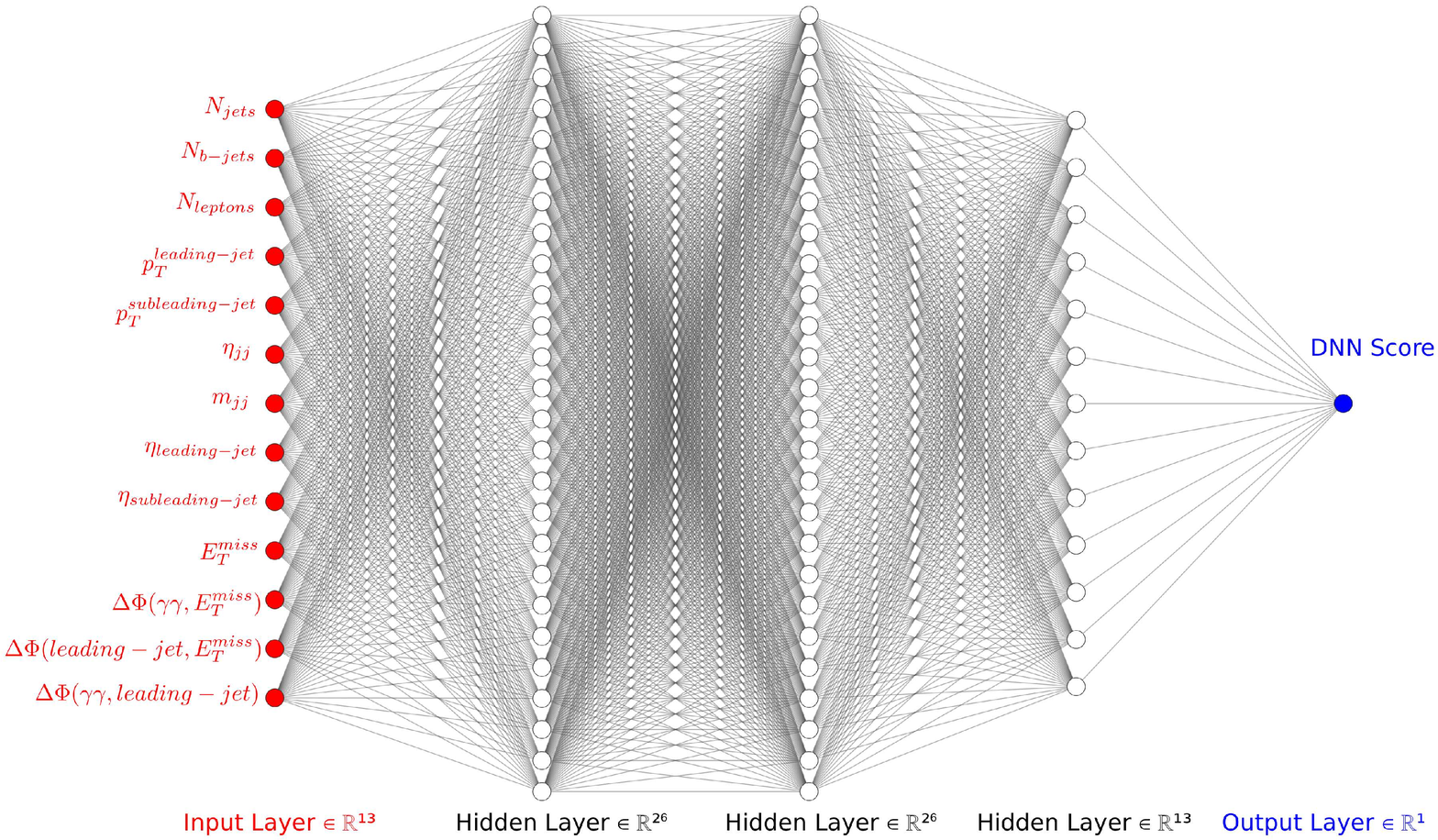}
\caption{DNN architecture for the classification of events as background or signal in this study, with the \texttt{"Layout=TANH|26,TANH|26,TANH|13,LINEAR" configuration.}\label{fig1:DNNTOPO}}
\end{figure}

Since we are dealing with a binary classification problem, where the events are classified as signal or background, the cross-entropy (see equation \ref{eqn:lossEntropy}) is used as a loss function. A normalised initialization of the weights, with zero-mean Gaussian distributed random values, is chosen according to Ref.~\citen{GlorotXavier}. These weights control the multiplicative effect of the different layers. Figure~\ref{fig1:DNNTOPO} describes the network topology and the input variables used in this study. Furthermore, in order to handle the problem of imbalances in training samples\cite{10.1007/978-3-540-27868-9_88}, the signal and background training events are renormalised such that their respective sums of effective weighted events are the same. The training of the classifier is divided into three stages. The first stage begins with a learning rate of $10^{-2}$ and a momentum of 0.9. In order to avoid overtraining the classifier, the weights are regularised using the $L2$ option, which multiplies a scaling factor of $10^{-3}$ to the norms of the weight matrices. For the second training stage, learning rate and momentum are reduced to $10^{-3}$ and 0.5, respectively. For the final training phase, the learning rate is set to $10^{-2}$ and the momentum to 0.3. Finally, the performance of the neural network in these three stages is evaluated on the test samples. 

\subsection{Full supervision classification}
\label{sec:FSLC}

The full supervision classification described in Section~\ref{sec:FSL} was applied to each individual SM Higgs boson production mechanism against the total background. The background is labelled with ``0" while the signal is labelled with ``1". The performance of the proposed neural network topology was tested. The results of the training and testing samples have been compared to evaluate if the model is overtrained. The experimental or theoretical uncertainties are not implemented in this analysis, which may lead to a minor degradation of the performance.

\begin{figure}[hbt!]
\centering
 \subfigure[ggF]{\includegraphics[width=0.45\linewidth]{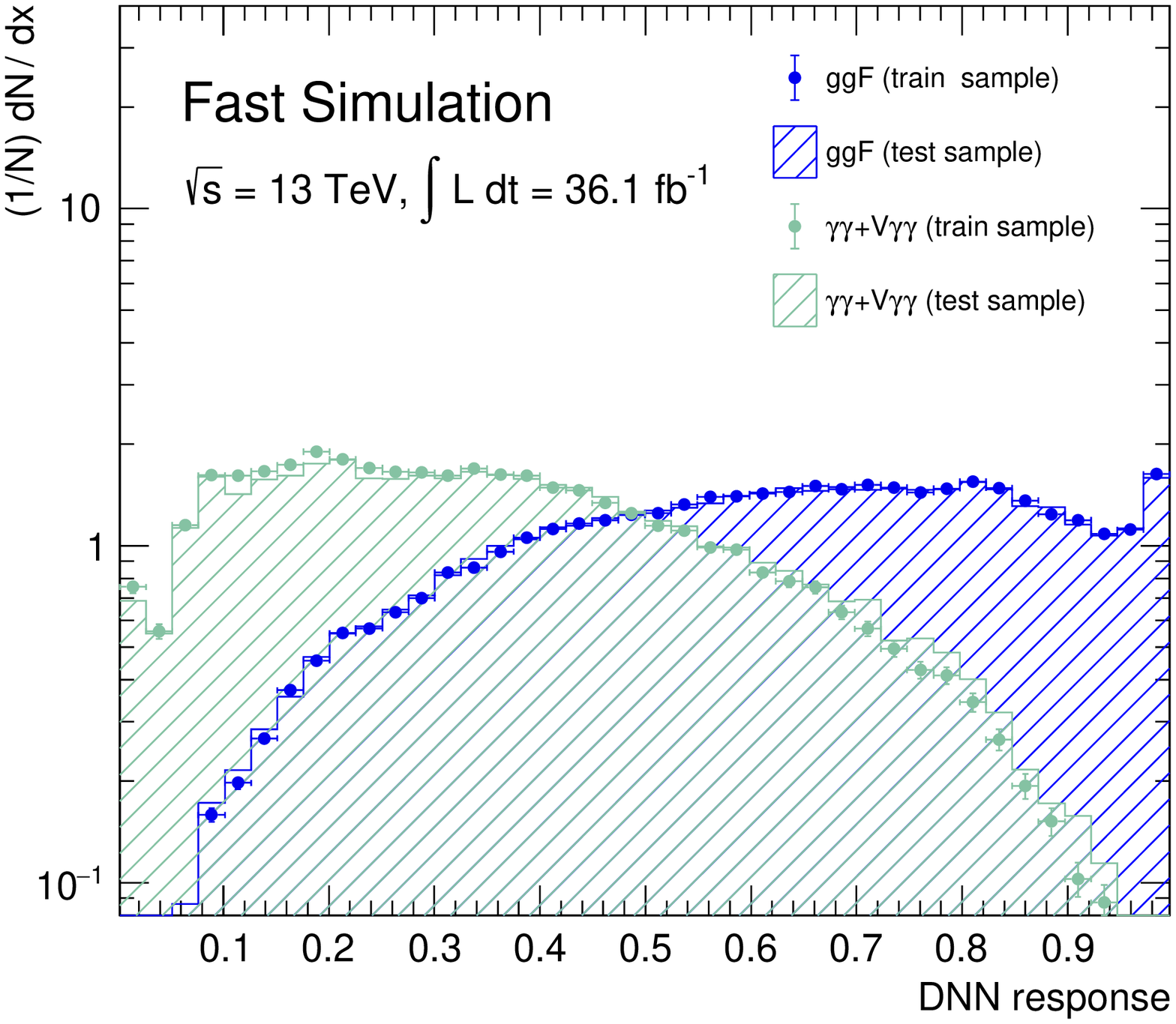}}
 \subfigure[VBF]{\includegraphics[width=0.45\linewidth]{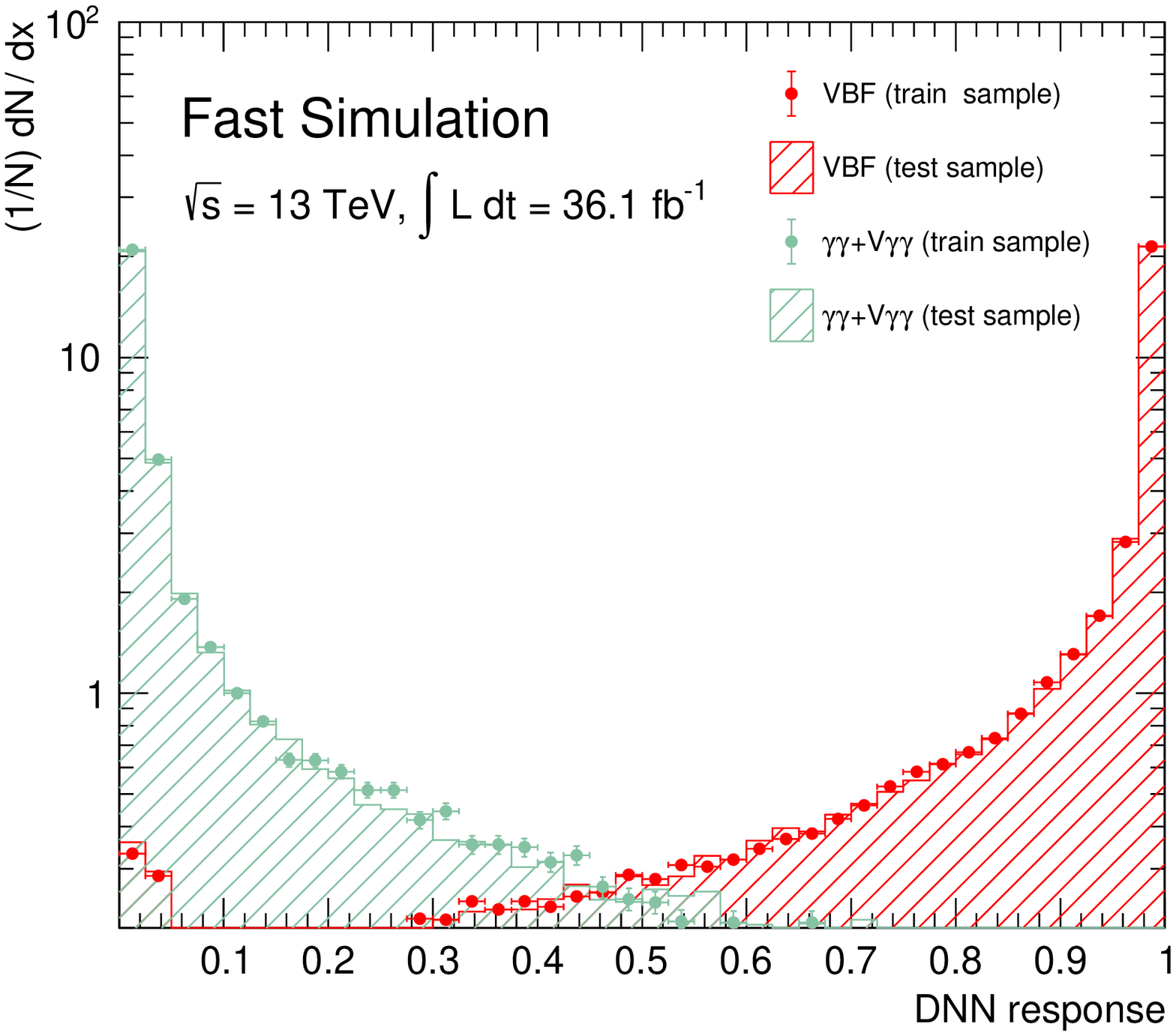}}
\caption{Full supervision DNN response distributions for the Higgs production, through: (a) gluon-gluon fusion and (b) VBF.}\label{fig1:FullDNNFus}
\end{figure}
\FloatBarrier

\begin{figure}[hbt!]
\centering
  \subfigure[$Wh$]{\includegraphics[width=0.45\linewidth]{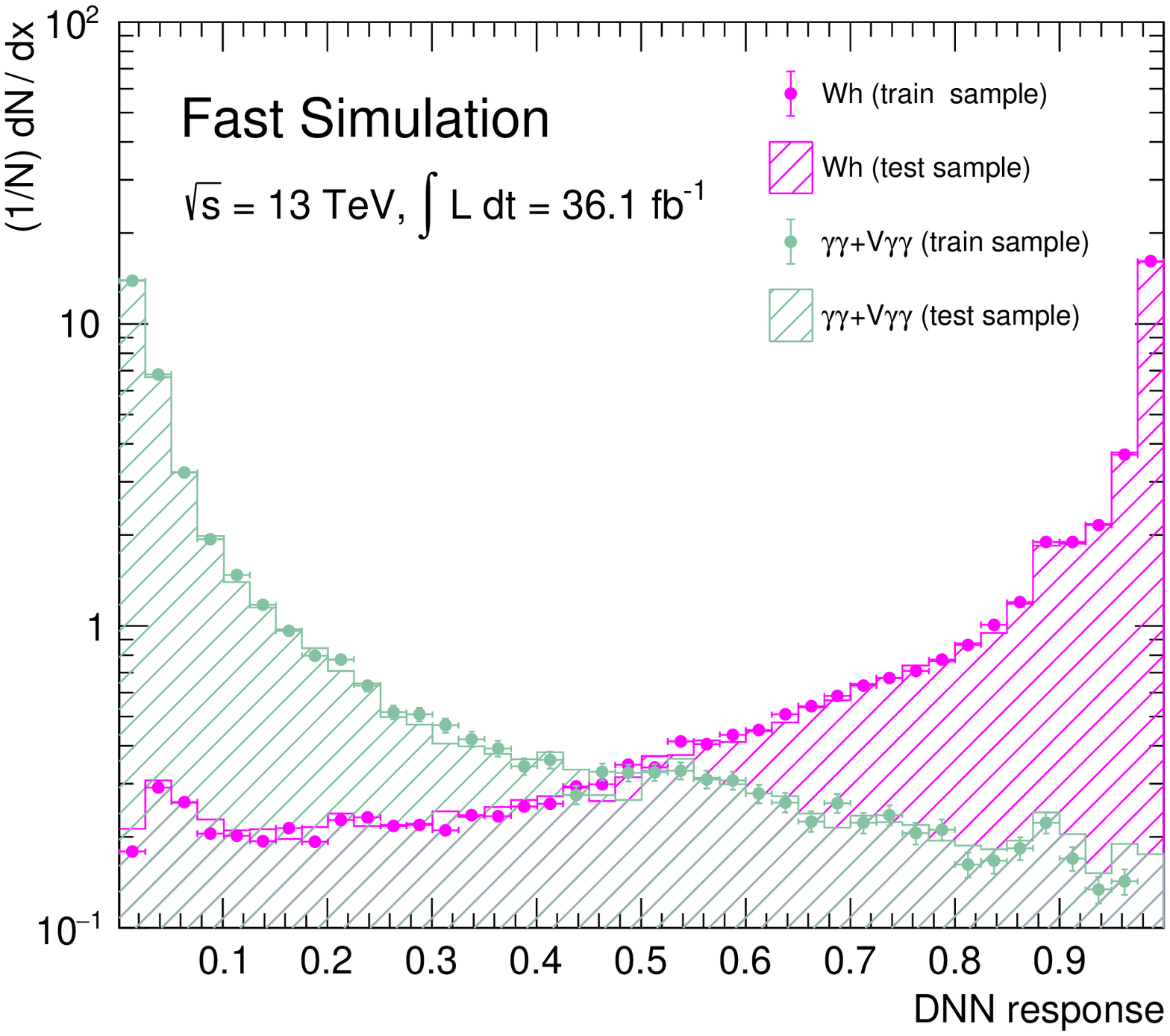}}
  \subfigure[$Zh$]{\includegraphics[width=0.45\linewidth]{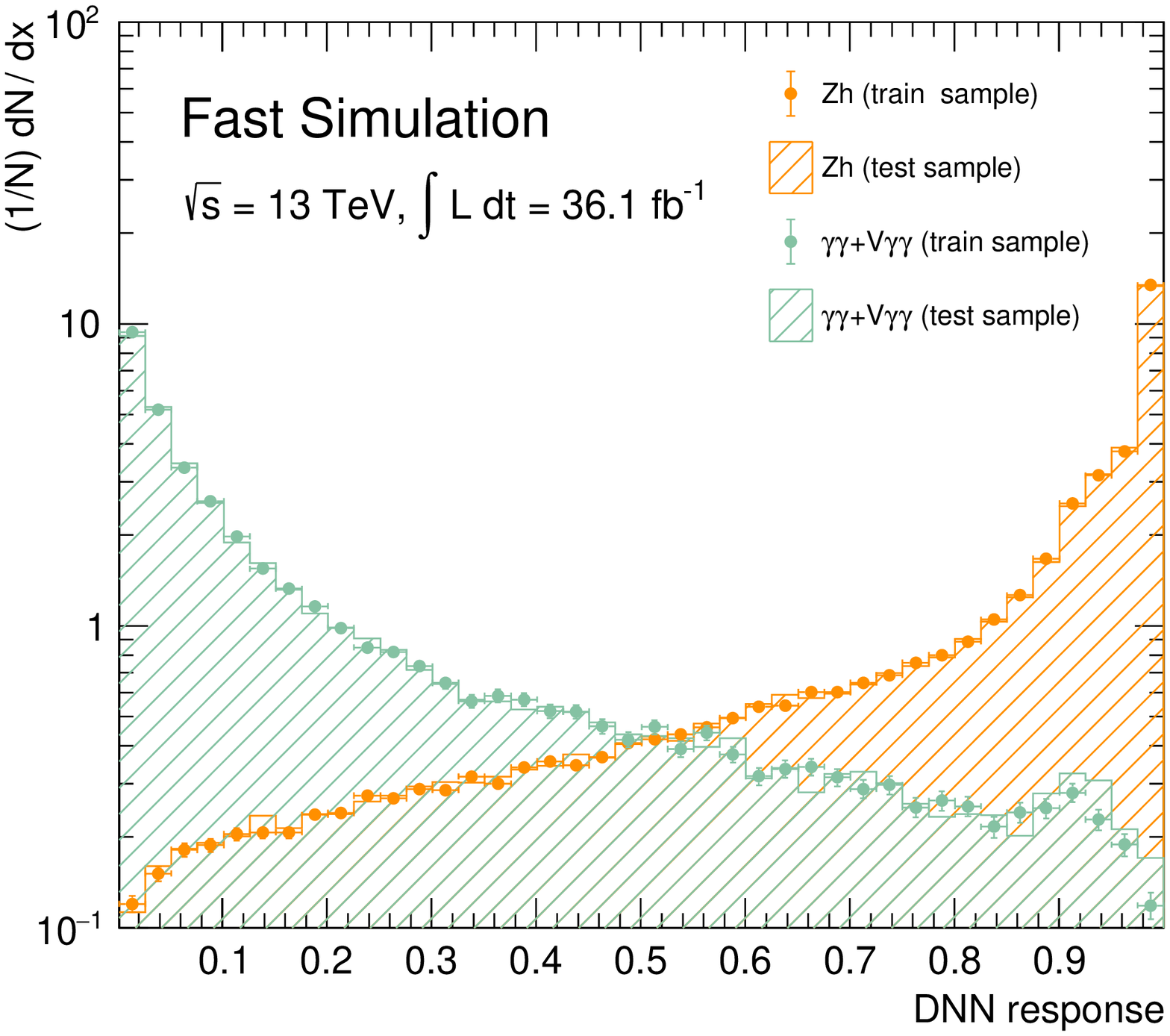}} \\
  \subfigure[$t\bar{t}h$]{\includegraphics[width=0.45\linewidth]{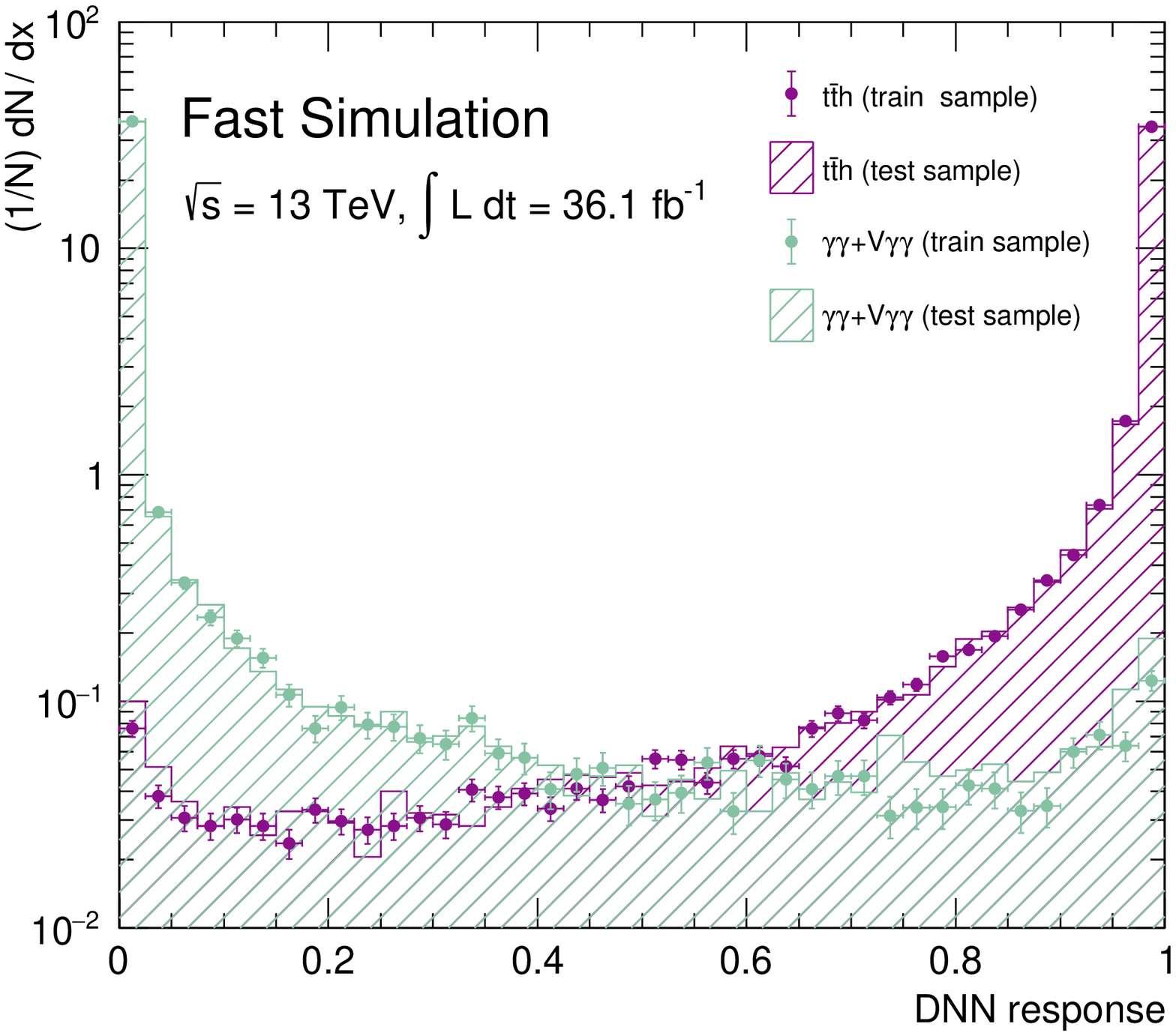}}
\caption{Full supervision DNN response distributions for the SM Higgs boson associated production, with: (a) $W$ vector boson, (b) $Z$ vector boson or (c) a top quark pair. }\label{fig1:FullDNNAsso}
\end{figure}
\FloatBarrier

Figures~\ref{fig1:FullDNNFus} and ~\ref{fig1:FullDNNAsso} show the fully supervised DNN response distributions for ggF with VBF and the associated productions, respectively. As expected, there is clear separation between signal and background processes.

\begin{figure}[hbt!]
\centering
\includegraphics[width=0.60\linewidth]{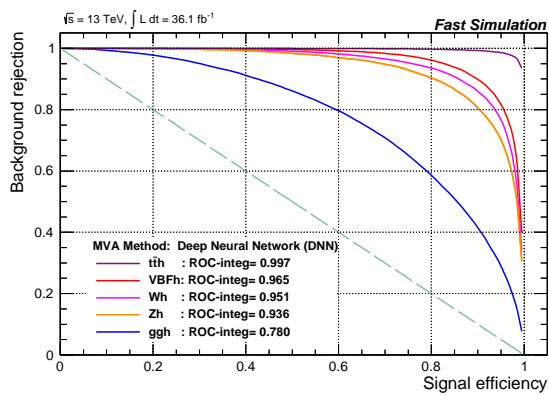}
\caption{ROC curves from full supervision DNN classification, for different Higgs production mechanisms.}\label{fig1:FullROC}
\end{figure}

The signal to background separation power of the DNN classifier is evaluated using the Receiver Operator Characteristic (ROC) curve, which quantifies how accurately the classifier is able to differentiate between signal and background. A performance metric is the Area Under the Curve (AUC) which measures the entire two-dimensional area under the ROC curve. Figure~\ref{fig1:FullROC} shows a plot of the background rejection versus signal efficiency for different Higgs production mechanisms. The ROC curves show hierarchical separation of ggF, VBF and associated productions ($t\overline{t}h$, $Zh$ and $Wh$). 

The separation power for each production mechanism is evident and dependent on the respective variable separation. The ROC integral of the ggF production mechanism is only 78\%, due to the fact that it populates the phase-space similarly to the background. The VBF and associated productions are similar, while the $t\bar{t}h$ stands out because of the multiplicity of signatures and topologies.

\FloatBarrier
\subsection{Supervised classification with unlabelled signals}
\begin{figure}[hbt!]
\centering 
  \includegraphics[width=0.45\linewidth]{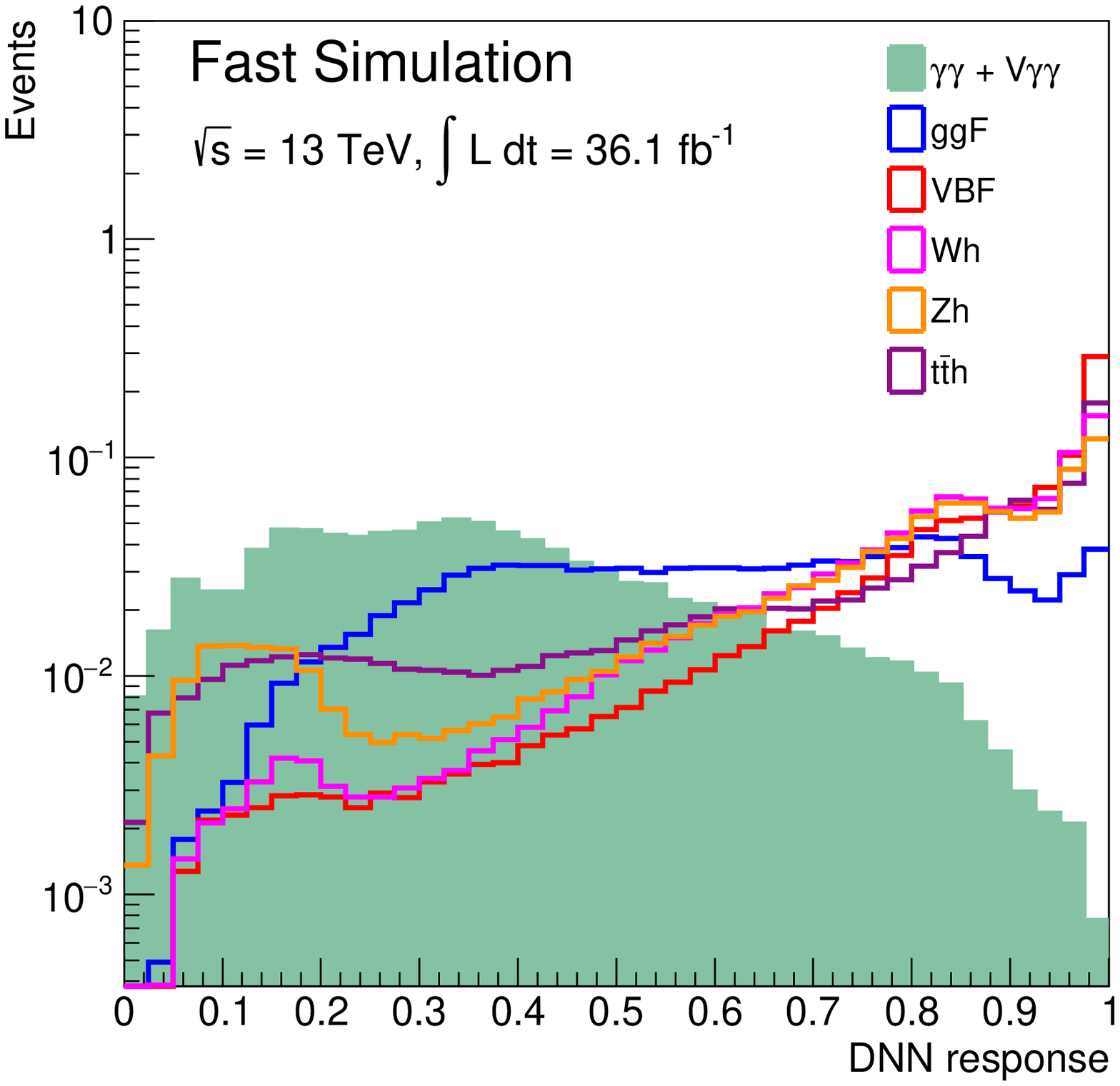}
  \includegraphics[width=0.45\linewidth]{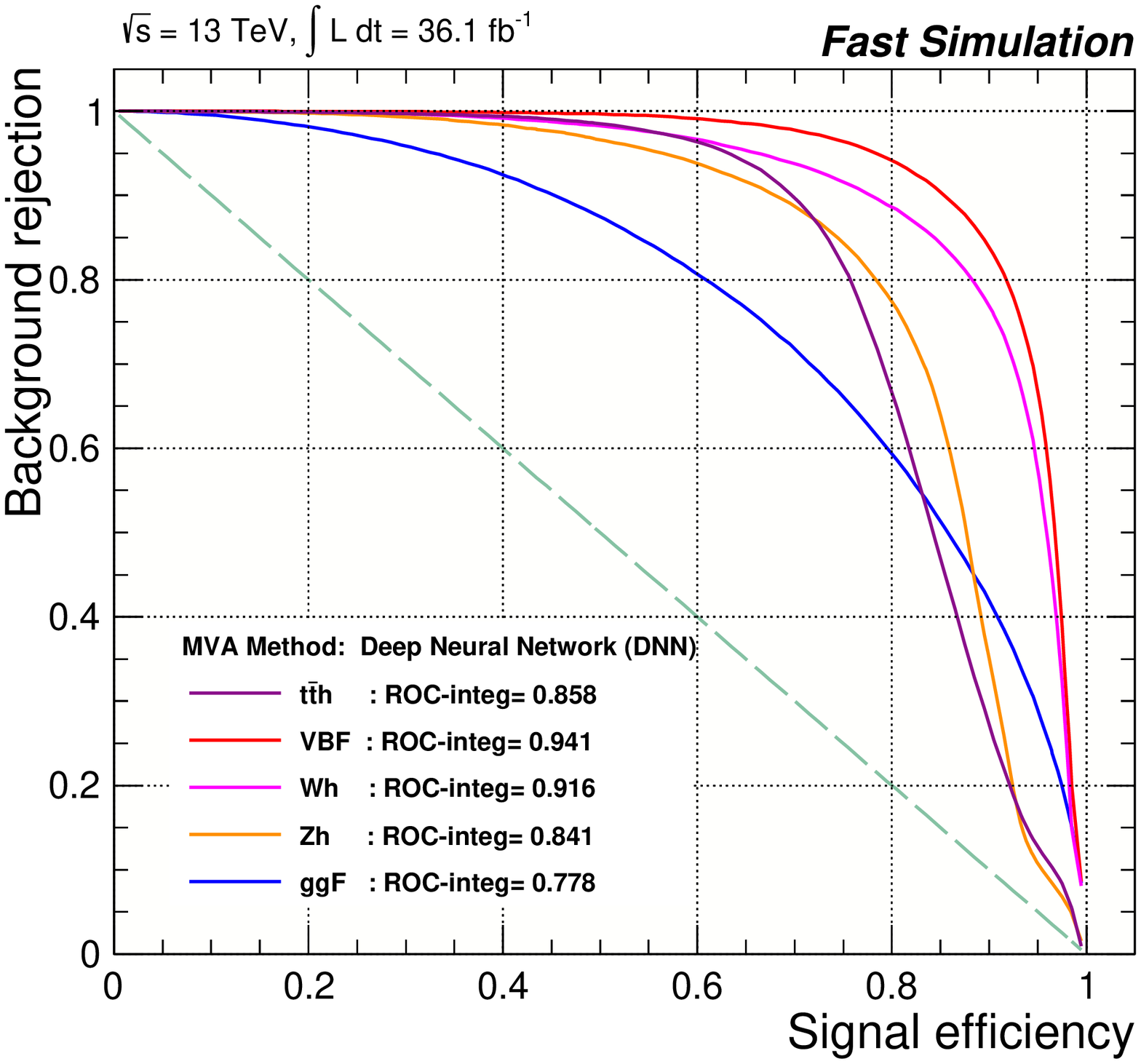}
\caption{Supervised learning results with unlabelled signals. Left: DNN response distributions. Right: ROC curves, for different Higgs production mode.}\label{fig1:UnlabelSuper}
\end{figure}

The performance of the supervised classification method with unlabelled signals is evaluated through its ability to distinguish between different SM Higgs boson production processes, as explained in Section~\ref{sec:UFSL}. A signal sample, consisting of a mixture of five production processes, is created and labelled as class ``1", while the background is labelled as class ``0".  Figure~\ref{fig1:UnlabelSuper} shows a generic scan of the DNN classifier output distribution of the individual Higgs productions and the total background. 
The ROC curves demonstrate that the separation power of the supervised learning method with unlabelled signal is noticeably worse than that of the fully labeled supervised learning. However, the response on ggF remains consistent since it is the dominant mechanism, constituting more than 85\% of the Higgs production. This observation is attributed to the different event yields for each production mechanism. In order to quantify the effects of each production mechanism's relative contribution on the classifier, the expected cross-sections are increased. This is done by normalising the VBF, $Wh$, $Zh$ and $t\bar{th}$ production mechanisms to different fractions of the expected ggF production yield.

Table~\ref{fig1:unlROCTable} summarises the results from~\ref{sec:Statist}, where the integral of the ROC curves for all the production mechanisms increase with respect to the ggF cross-section fraction. 

\FloatBarrier

\begin{table}[hbt!]
\tbl{Relative contribution effect of each signal on the performance of the DNN classifier, for supervised learning with unlabelled signal (\ref{sec:Statist}). The signals are normalised to different fractions of the expected events from the ggF production mechanism. }
{\begin{tabular}{l|ccccc}
\hline
\hline
 & \multicolumn{5}{c}{ROC integral normalised to}\\
Process & $\sigma_{process}$ & 0.25 $\sigma_{ggF}$ &  0.5 $\sigma_{ggF}$ & 0.75 $\sigma_{ggF}$ & $\sigma_{ggF}$ \\
\hline
\hline
ggF  & 0.778 & - & - & - & - \\
VBF & 0.941 & 0.946 & 0.951 & 0.954 & 0.961 \\
$Wh$ & 0.916 & 0.934 & 0.941 & 0.943 & 0.944\\
$Zh$ & 0.841 & 0.892 & 0.916 & 0.920 & 0.920\\
$t\bar{t}$h & 0.858 & 0.975 & 0.984 & 0.986 & 0.987 \\
  \hline
\hline
\end{tabular}\label{fig1:unlROCTable}}
\end{table}
\FloatBarrier
\subsection{Weak supervision classification}
\label{sec:WSL}

\begin{figure}[t]
\centering
  \includegraphics[width=0.60\linewidth]{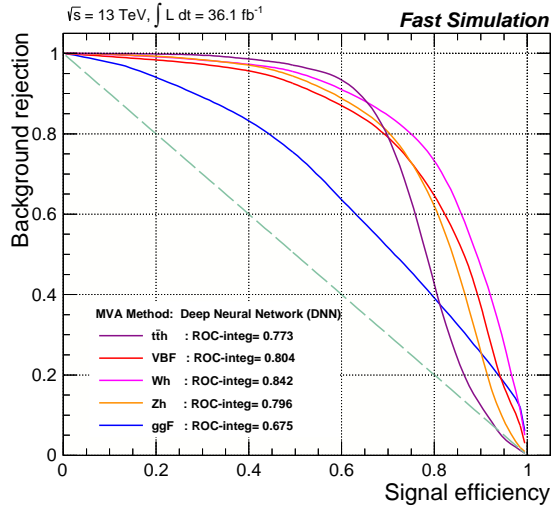}
\caption{ROC curves from weak supervision learning, using DNN classifier for different Higgs production mechanisms. Signals and background are normalised to its cross-sections.}\label{fig1:WeakROC}
\end{figure}

For the weak supervised learning study on the Higgs production mechanisms, the signals and background processes are mixed in the di-photon invariant signal mass window of 120\,GeV to 130\,GeV. The sideband contains only pure background in the region between 105\,GeV and 160\,GeV, excluding the signal mass window. The purpose of this procedure is to evaluate the ML algorithm's ability to classify a signal from the signal mass window of an unknown mixture of signal and background, with respect to pure background in the sideband. According to the ROC analysis results shown in Figure~\ref{fig1:WeakROC}, it is evident that weak supervision is capable of separating different signal candidates. However, the signal and background classification is not satisfactory when compared with the full supervision and full supervision with unlabelled signals, as summarised in Table~\ref{table2:WeakROCTable}. Given these points, weak supervision seems to be inefficient due the complexity of the phase-space explored here. For that reason a guided weak supervision method with topological requirements is introduced. This will be discussed explicitly in Section~\ref{sec:WSLTopo}. 

\begin{table}[t]
\tbl{Comparison of the integral of the ROC curve response from different supervised learning methods.}
{\begin{tabular}{l|ccc}
\hline
\hline
Process & Full Supervision & Unlabelled Full Supervision & Weak Supervision \\
\hline
\hline
ggF & 0.780 & 0.778 & 0.675 \\
VBF & 0.965 & 0.941 & 0.804 \\
$Wh$ & 0.951 & 0.934 & 0.842\\
$Zh$ & 0.936 & 0.841 & 0.796\\
$t\bar{t}$h & 0.997 & 0.858 & 0.773\\
  \hline
\hline
\end{tabular}\label{table2:WeakROCTable}}
\end{table}

\section{Weak supervision learning with topological requirements}
\label{sec:WSLTopo}

The performance of supervised learning methods on the classification of SM Higgs boson production modes has been evaluated and compared above. The results have demonstrated that the performance of the fully supervised learning is superior to that of unlabelled and weak supervised learning. However, it is worth noting that the performance of the full supervision and unlabelled supervision methods are comparable. The discrepancies are attributed to the relative contributions from each of the SM Higgs boson production mechanisms. Taking for instance the ggF, which has the largest cross section at the LHC and almost the same topology as the main background, a ROC integral difference of 0.2$\%$ is observed (see Table.~\ref{table2:WeakROCTable}). For the $t\bar{t}h$ production mechanism, which only contributes less than $1\%$ of the total production cross-section, the ROC integral differential is 14$\%$. The weak supervised learning method remains the worst performing. Therefore there is a need to implement the method that we refer to as the guided weak supervision method. In guided weak supervision, the training is restricted to the topological signatures derived from selected production mechanisms. This methodology will be evaluated and verified with various SM Higgs boson production mechanisms.

The SM Higgs boson production mode through VBF represents the second largest cross section at the LHC. The VBF mechanism in the SM entails the scattering of two quarks, leading to two hard jets in the forward and backward regions of the ATLAS detector with high di-jet invariant mass, $m_{jj}$. These topological characteristics can be exploited by the DNN to distinguish VBF from backgrounds and other production mechanisms. In this context, the VBF topology is defined as events with at least two jets. This is followed by a signal significance scan using $m_{jj}$ and the pseudorapidity separation between the two leading jets, $\Delta\eta_{jj}$. We use the signal significance scan to maximise the relative contribution of the VBF production mechanism and minimise the background yield as much as possible. 

\begin{figure}[t]
\centering
  \includegraphics[width=0.45\linewidth]{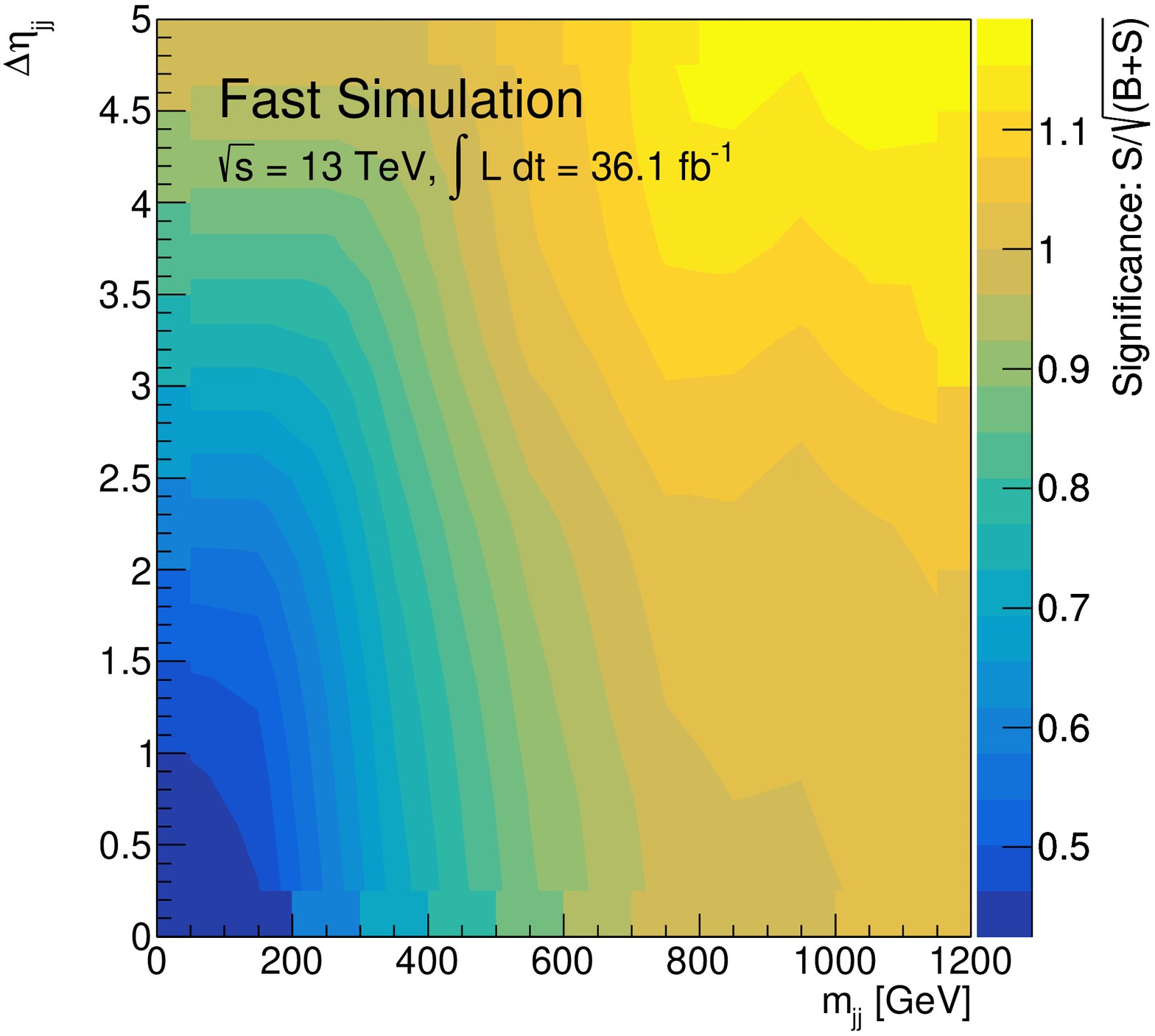}
  \includegraphics[width=0.45\linewidth]{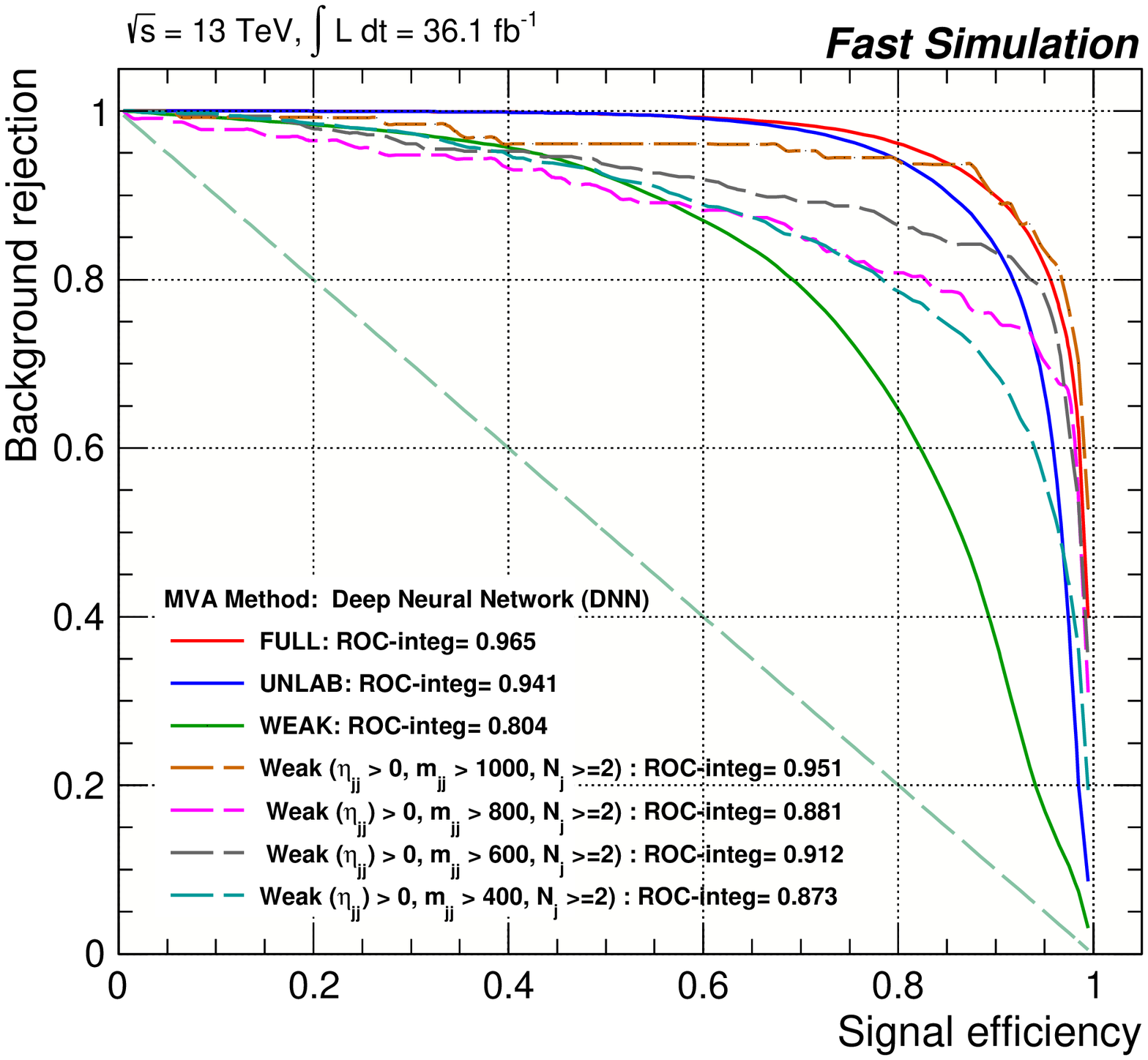}
\caption{Weak supervision learning results with VBF topological selections. The graph on the left shows two-dimensional scan of signal significance with respect to  $m_{jj}$ and $\Delta\eta_{jj}$ for the VBF production mode. The $z$-axis corresponds to the expected signal significance. The graph on the right displays the ROC curves for the VBF production mechanism with a variety of different learning methods (see text).}\label{fig1:ROCTOPODNN}
\end{figure}
\FloatBarrier

\begin{figure}[hbt!]
\centering
  \includegraphics[width=0.6\linewidth]{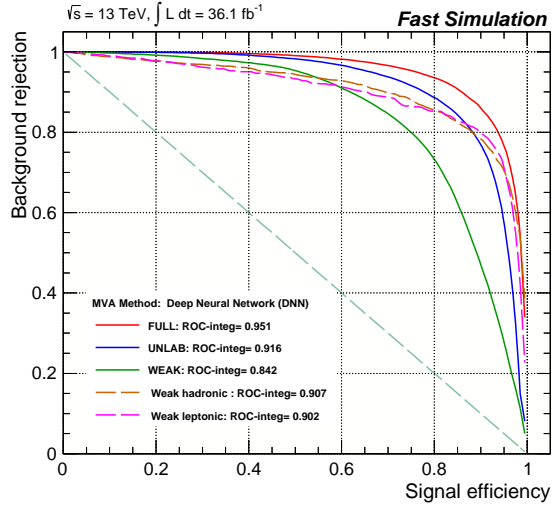} 
\caption{ROC curves for a variety of different learning methods in inclusive $Wh$ final states are represented with solid lines. The weak supervision learning with leptonic and hadronic $Wh$ topological requirements are shown with dashed lines.}\label{fig1:ROCTOPODNNwh}
\end{figure}

\begin{figure}[hbt!]
\centering
  \includegraphics[width=0.6\linewidth]{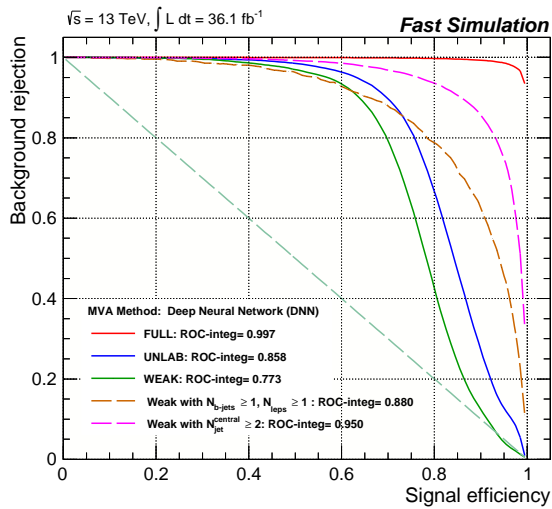} 
\caption{ROC curves for a variety of different learning methods in inclusive $t\bar{t}h$ final states are represented with solid lines. The weak supervision learning with $t\bar{t}h$ topological requirements are shown with dashed lines.}\label{fig10:ROCTOPODNNtth}
\end{figure}

Figure~\ref{fig1:ROCTOPODNN} shows a ROC curve comparison between full supervision, full supervision with unlabelled signals and weak supervision at the pre-selection level i.e. without applying the VBF selection. The weak supervised learning methods, with topological requirements, are highlighted with dashed lines. The training of the neural networks is performed in four regions determined with respect to the signal significance. The results reveal various improvements in the weak supervised classification with topological requirements. This attests to the capability of separating the VBF from the other production processes, thus qualifying it to be comparable with full supervised learning.  
A similar test is performed on the associated production of Higgs boson, with a $W$ gauge boson, where the neural networks are trained separately with two topological features from the $Wh$ final states. This includes leptonic and hadronic decays of the $W$ boson. Figure~\ref{fig1:ROCTOPODNNwh} shows the performance response of full supervision, full supervision with unlabelled signals and weak supervision for the inclusive $Wh$ final states, as well as the leptonic and hadronic decays. The performance of weak supervision classification also experiences improvements with topological requirements and is able to isolate the $Wh$ process.
In addition, the guided weak supervision is tested on the production of the Higgs boson in association with top-antitop quarks. Since the relative contribution of the top-antitop quark to the total Higgs boson production is less than 1\%, it becomes challenging to extract this signal process. Most prominently, this is found to be the case for weak supervision. However, the test is performed by requiring two topological characteristics, representing a fully hadronic (at least two central jets) and semi-leptonic (at least one $b$-jets and one lepton) decays of top quark from $t\bar{t}h$ process. The ROC curves of the guided weak supervision for the two typologies are highlighted in Figure~\ref{fig10:ROCTOPODNNtth}. The ROC integral values are an illustration of the ability of the methodology to extract small signals. Finally, the effectiveness of guided weak supervision for the VBF, $Wh$ and $t\bar{t}h$ production mechanisms is quantified and compared to the inclusive weak supervision in Table~\ref{table3:WeakROCTopoSumm}.
\begin{table}[t]
\tbl{Summary of the ROC integrals for VBF, $Wh$ and $t\bar{t}h$ with different topological requirements, compared to inclusive weak supervision learning.}
{\begin{tabular}{llc}
    \hline
    \hline
    Process & Weak Supervision & ROC Integral \\
    \hline
    \hline
    \multirow{5}{*}{VBF}& Inclusive & 0.804\\
    & $N_{jets} \geq 2$, $\Delta\eta_{jj}$ > 0, $m_{jj}$ > 1000\,GeV & 0.951\\
    & $N_{jets} \geq 2$, $\Delta\eta_{jj}$ > 0, $m_{jj}$ > 800\,GeV & 0.881\\
    & $N_{jets} \geq 2$, $\Delta\eta_{jj}$ > 0, $m_{jj}$ > 600\,GeV & 0.912\\
    & $N_{jets} \geq 2$, $\Delta\eta_{jj}$ > 0, $m_{jj}$ > 400\,GeV & 0.873\\
    \hline
    \hline
    \multirow{3}{*}{$Wh$}& Inclusive & 0.842\\
    & Hadronic & 0.907\\
    & Leptonic & 0.902\\
    \hline
    \hline
    \multirow{3}{*}{$t\bar{t}h$}& Inclusive & 0.773\\
    & $N_{b-jets} \geq 1$, $N_{leptons}  \geq 1$ & 0.880\\
    & $N_{b-jets} \geq 2$ & 0.950\\
    \hline
    \hline
\end{tabular}\label{table3:WeakROCTopoSumm}}
\end{table}
\FloatBarrier

\section{Discussion and conclusions}
\label{sec:Concl}

Searches for new resonances at the EW scale are hampered by relatively large SM backgrounds. Searches using inclusive and model dependent approaches have not identified corners of the phase-space with potential excesses over the SM-only hypothesis. This could indicate that new resonances may be produced in subtle topologies that standard searches have difficulty isolating. 

Full supervision is widely used for searches in HEP, which assumes detailed knowledge of the BSM signal. As a result, optimizations obtained on the basis of full supervision may lead to undesired biases. Searches for bosonic resonances produced in VBF can serve as an illustration of the bias of full supervision. A minor change in the CP structure of the coupling of the boson to weak bosons induces significant changes in the topology of the final state. When the search is optimised with SM-like couplings, a potential signal can be missed altogether. In order for the search to gain generality, it would need to be performed for a scan of model parameters. In general, this exercise can become cumbersome and is not performed in practice.

Machine learning may play a significant role in the deeper exploration of the phase-space available at the LHC. In order to evaluate the performance of different machine learning approaches, the SM Higgs boson with the di-photon decay is used. This allows us to understand the performance of machine learning in a situation where the particle is produced through different production mechanisms, thus populating various corners of the phase-space. The performance of the full supervision approach was compared to the unlabelled full supervision. The latter performs full supervision on a signal sample for which the different production mechanisms are not labelled. The performance of these two approaches is compared with that of weak supervision performed with an unlabelled signal sample. The performance of weak supervision is significantly worse (see Table~\ref{table2:WeakROCTable}) compared to that of full supervision for all the production mechanisms considered here. The effect is strongest for the $t\overline{t}h$ mechanism, which is the one that provides the largest amount of distinct signatures. The VBF mechanism, which provides a distinct topology, is the second most affected. This seems to indicate that weak supervision alone is not particularly efficient in disentangling signal and background in the presence of signatures and topologies. It is important to note that hyper-parameter optimizations have been performed in order to ensure that the results obtained here are not induced by a particular choice of ML parameters.

In order to alleviate the effect, we introduce signatures and loose topological requirements before the implementation of weak supervised training. The impact on the signal efficiency and background rejection, in this approach, is evaluated for the different production mechanisms, where significant improvements are observed with respect to the implementation of weak supervision inclusively. This is referred to as guided weak supervision. While weak supervision, as setup here, has the advantage of not relying on a model for the signal it is necessary to restrict the phase-space where the sideband and the signal region are confronted with each other. While signature and topological requirements are driven by physics considerations, the search is not biased by the phenomenology of a model with a particular set of parameters. 

The performance of the approach has been evaluated with the SM Higgs boson. The signatures and topologies used here in principle can be used in the search for heavier or lighter SM Higgs-like bosons. Provided that the topological requirements are loose enough, these can be used for more generic searches of bosons. Other signatures can be added, such as MSSM-inspired. For instance, it is well motivated to apply weak supervision in the presence of a $b$-tagged jet, while requiring a maximum amount of jets in order to remove potential signals in association with top quarks. This can be extended to other types of searches, such as search for resonances in multi-particle cascades. 

It is important to note that the exploration of the phase-space remains driven by physics considerations. This is in contrast to general searches. That said, the concept of signal optimization and expected significance that is inherent to searches based on full supervision, needs to be revised and to be adapted to this approach. Emphasis needs to be made on model coverage in terms of the ability of the search to cover wide corners of the parameter space. 

Weak supervision has been defined here in the context of a search that uses sidebands to model the background. In principle, this approach can be applied to control samples in general. It can also be applied to signatures that don't display narrow resonances.

\section*{Acknowledgments}
The authors are grateful for support from the South African Department of Science and Innovation through the SA-CERN program and the National Research Foundation for various forms of support. The authors are also indebted to the Research Office of the University of the Witwatersrand for grant support.

\appendix
\section{Kinematic features}
\label{sec:inputVar}
\begin{figure}[hbt!]
\centering
\subfigure[]{\includegraphics[width=0.4\linewidth]{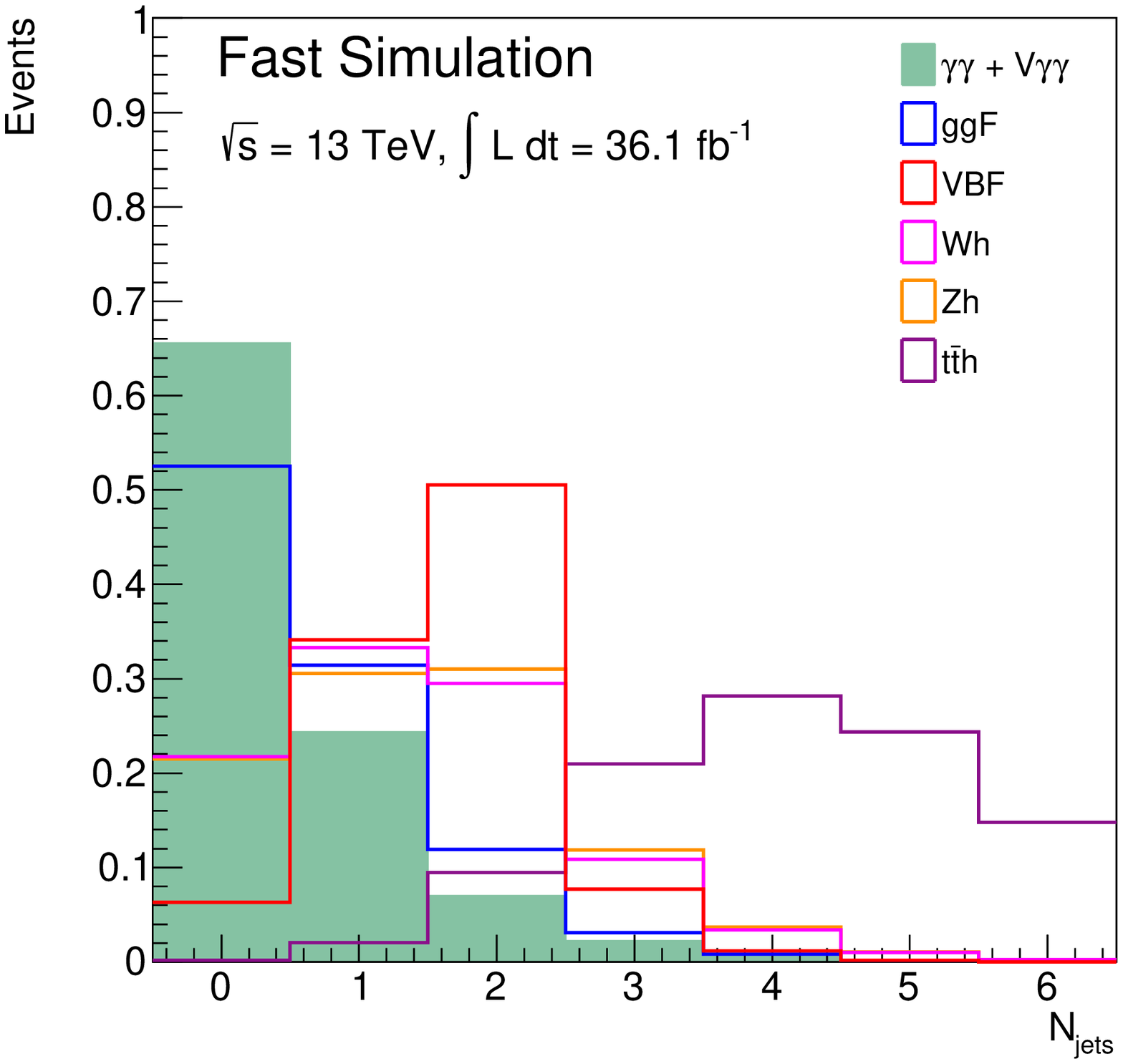}}
\subfigure[]{\includegraphics[width=0.4\linewidth]{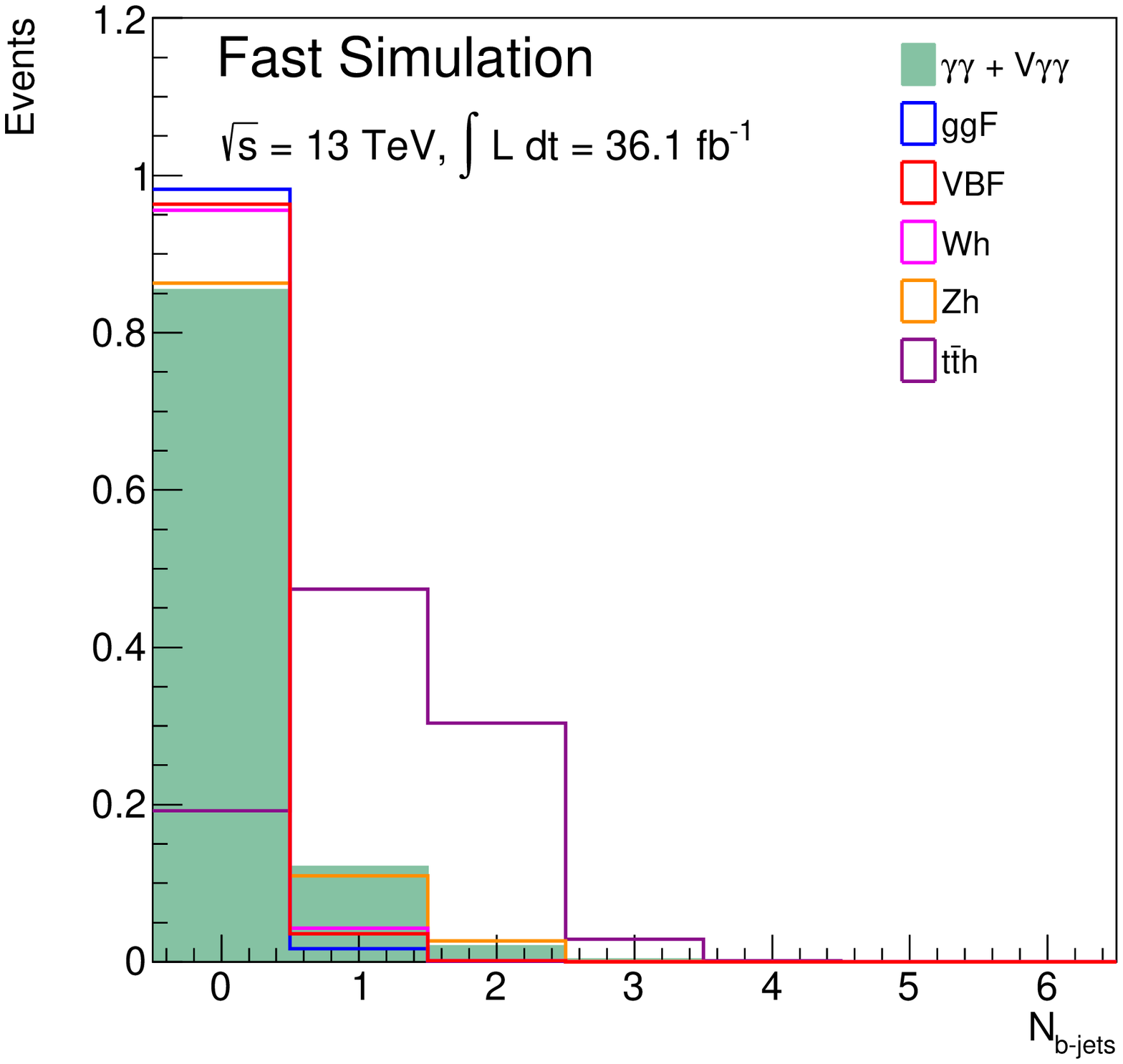}} \\
\subfigure[]{\includegraphics[width=0.4\linewidth]{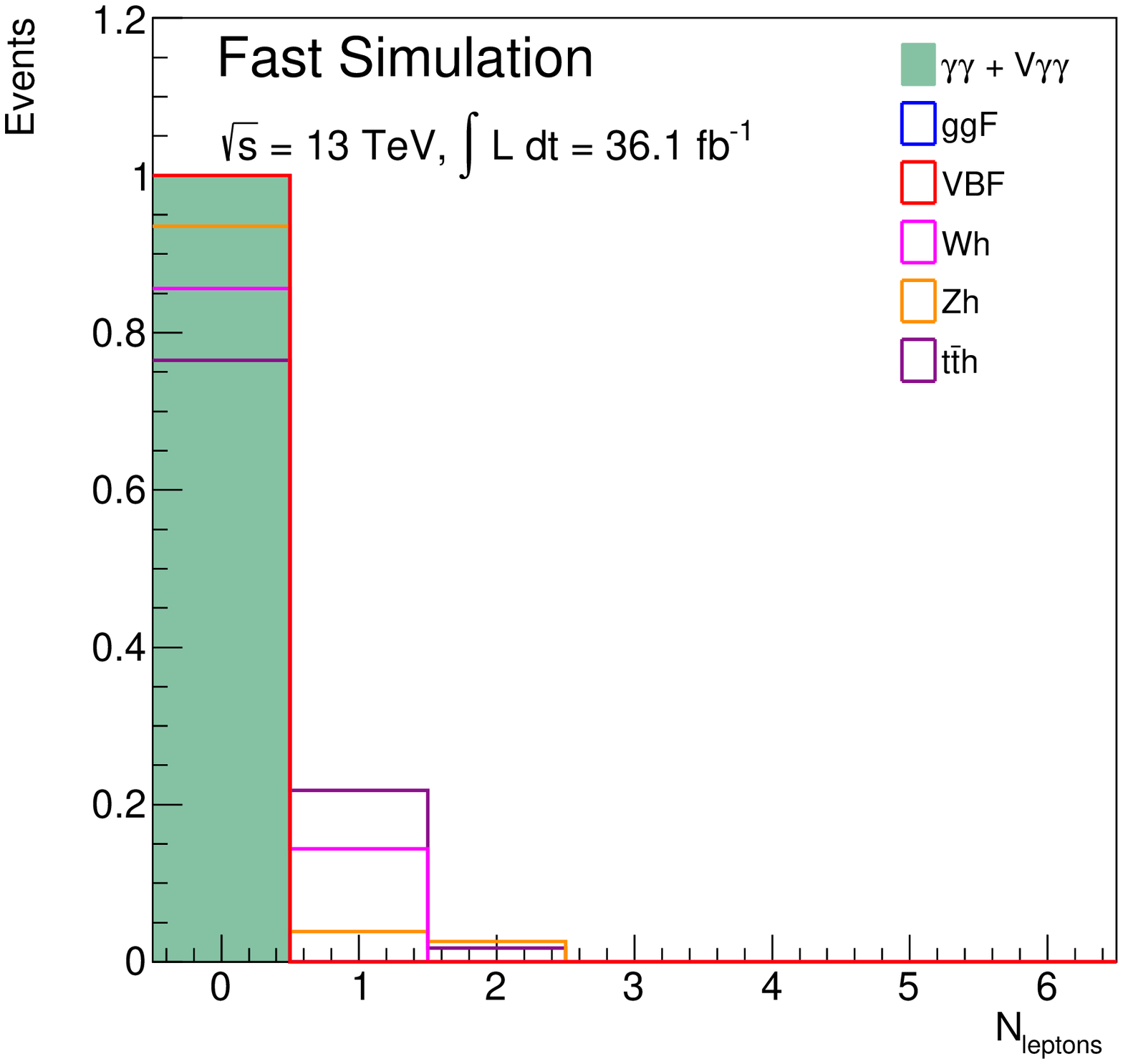}}
\caption{Signal and background distributions of the objects multiplicity: (a) number of reconstructed jets per event, (b) number of $b$-jets per event as identified with $b$-tagging algorithms and (c) number of the reconstructed leptons per event.}\label{fig1:Input1}
\end{figure}

Particle physics features implemented in machine learning represent input quantities that describe objects in the ATLAS detector. In this analysis a total of 13 discriminating features are used to classify background and different SM Higgs boson production mechanisms. They are chosen to exploit the phase space associated with each process. These input variables include information related to the object multiplicity, such as the number of reconstructed jets, number of $b$-jets (identified using $b$-tagging algorithms) and the number of the reconstructed leptons (see Fig.~\ref{fig1:Input1}). In addition, these inputs were used to assist the classifier to distinguish between processes with hadronic and leptonic decays, as well as to isolate mechanisms which lead to the presence of at least a $b$-jet in the final states, such as $t\bar{t}h$.

\FloatBarrier

\begin{figure}[hbt!]
\centering
\subfigure[]{\includegraphics[width=0.4\linewidth]{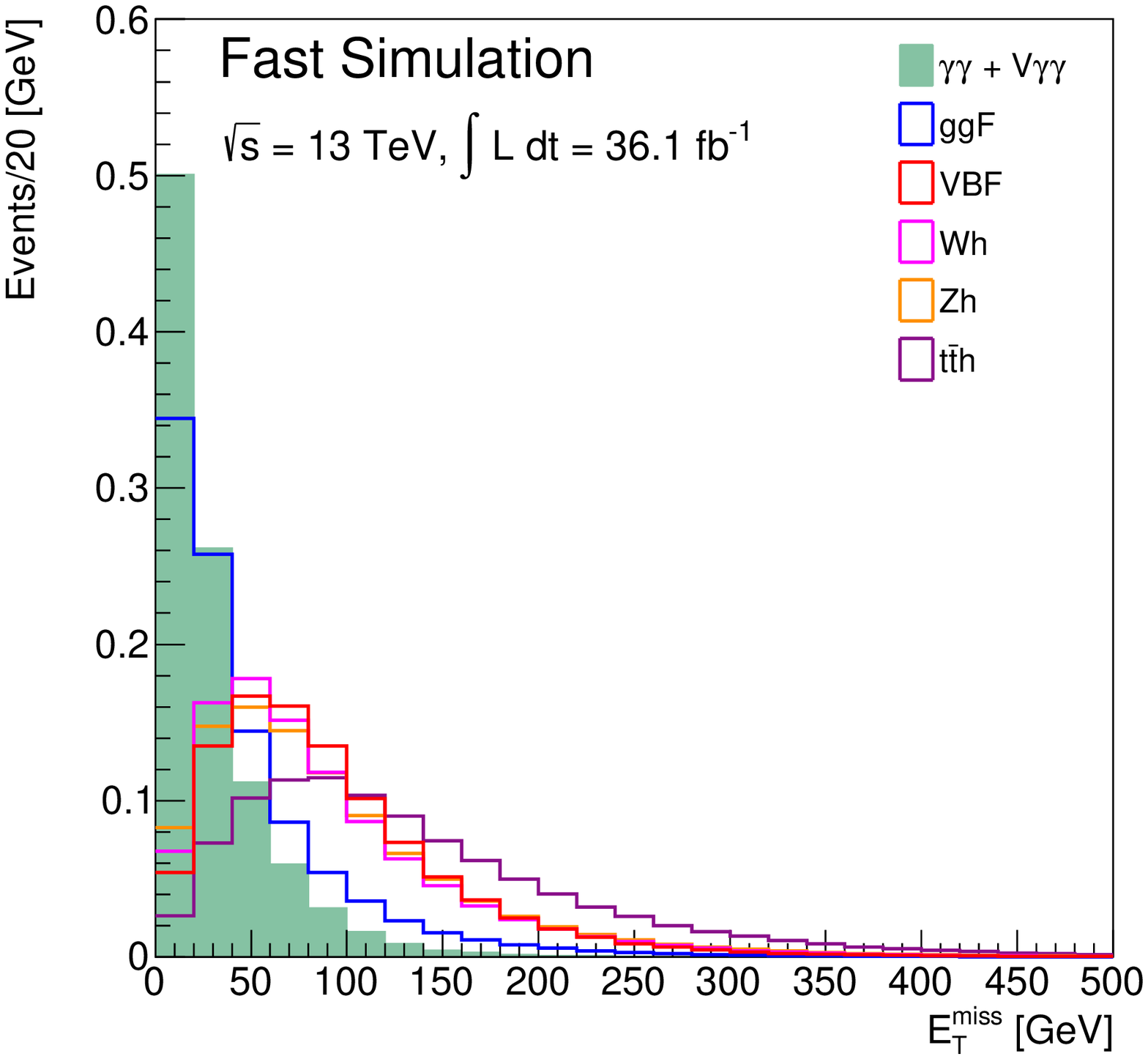}}
\subfigure[]{\includegraphics[width=0.4\linewidth]{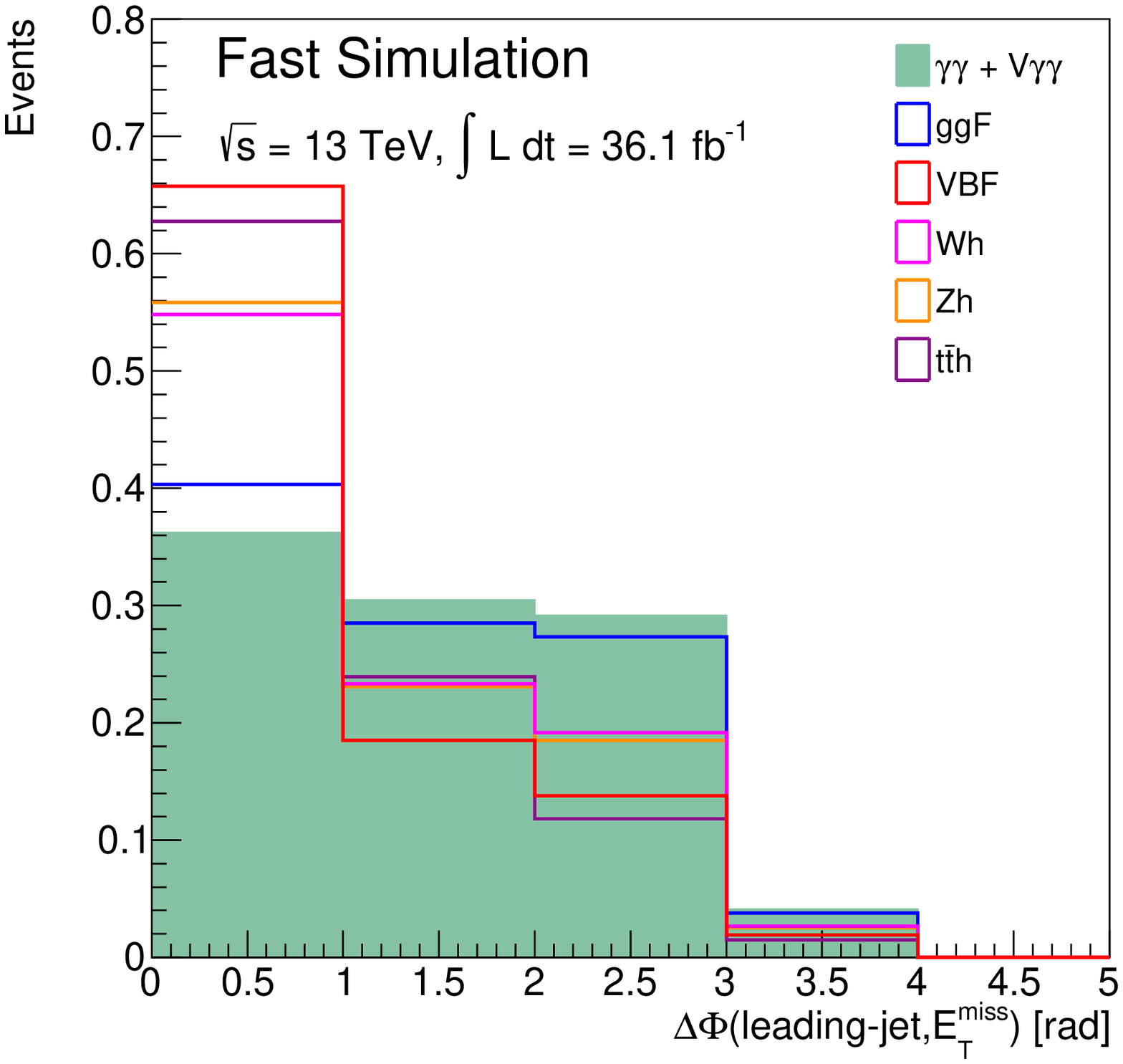}} \\
\subfigure[]{\includegraphics[width=0.4\linewidth]{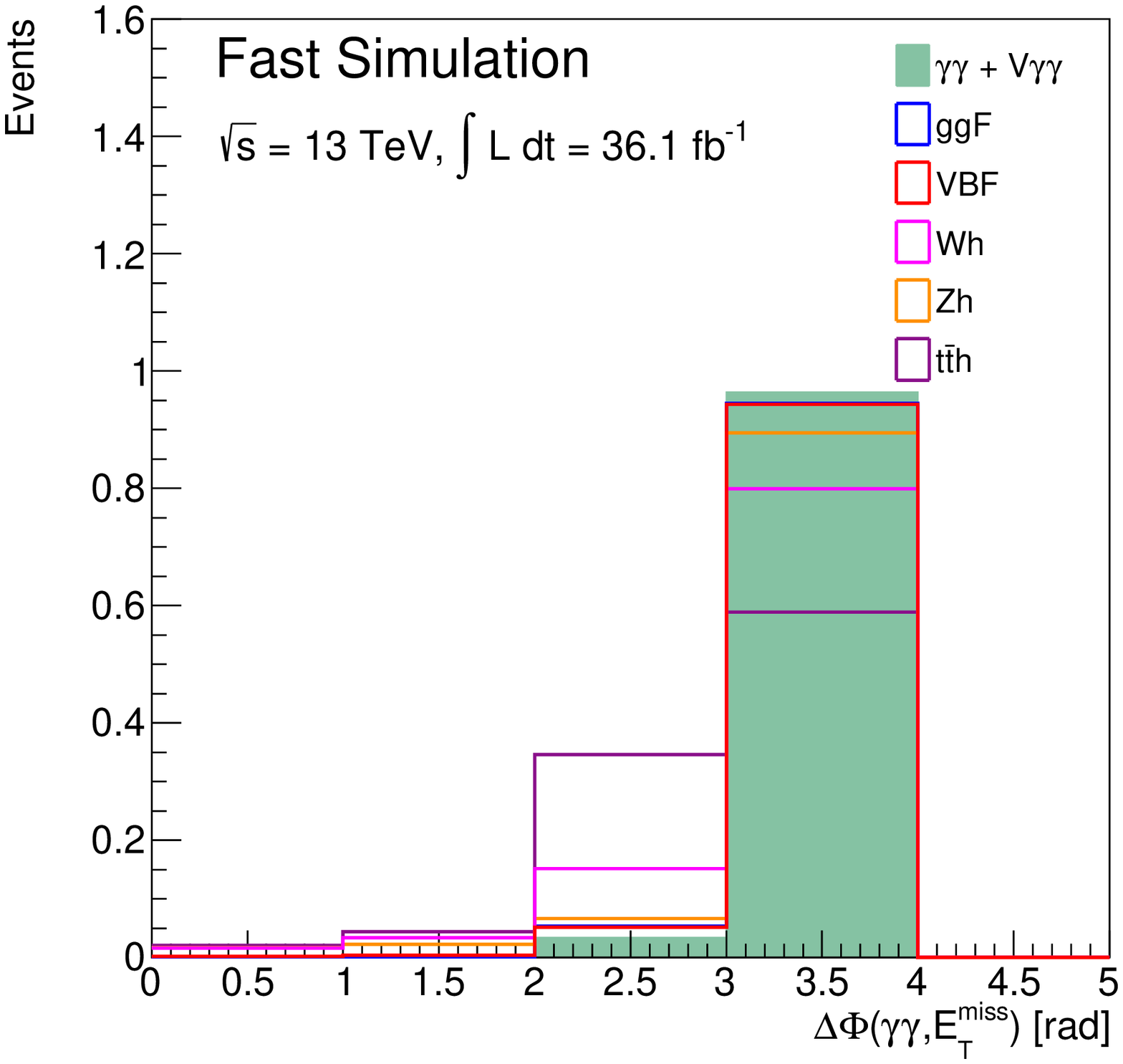}}
\subfigure[]{\includegraphics[width=0.4\linewidth]{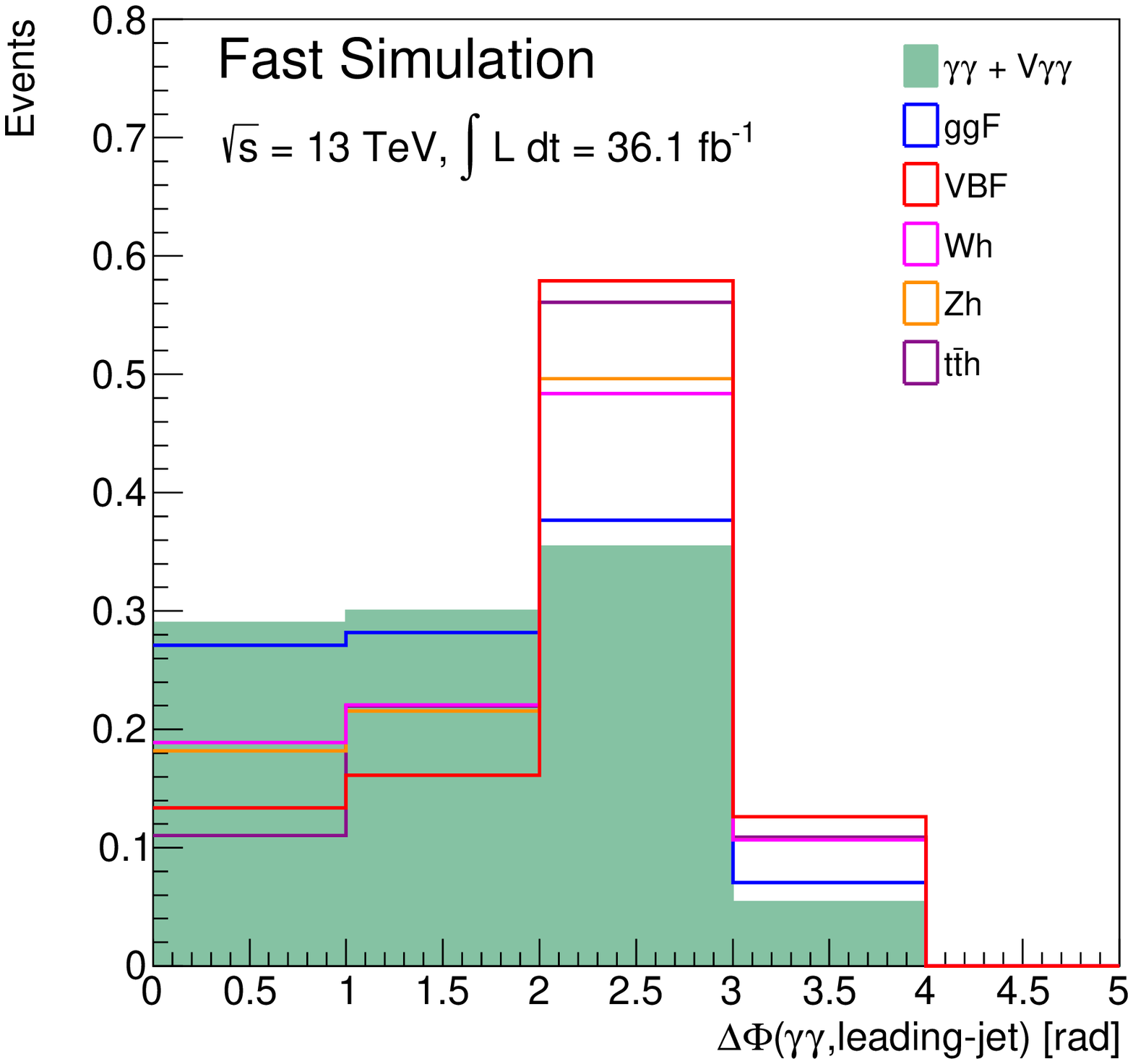}}
\caption{Signal and background distributions of the: (a) reconstructed missing transverse energy, (b) Azimuthal angle separation between the leading jets and the missing transverse energy, (c) Azimuthal angle separation between the di-photon system and the missing transverse energy (d) Azimuthal angle separation between the di-photon system and the leading jets.}\label{fig1:Input3}
\end{figure}

\begin{figure}[hbt!]
\centering
\subfigure[]{\includegraphics[width=0.4\linewidth]{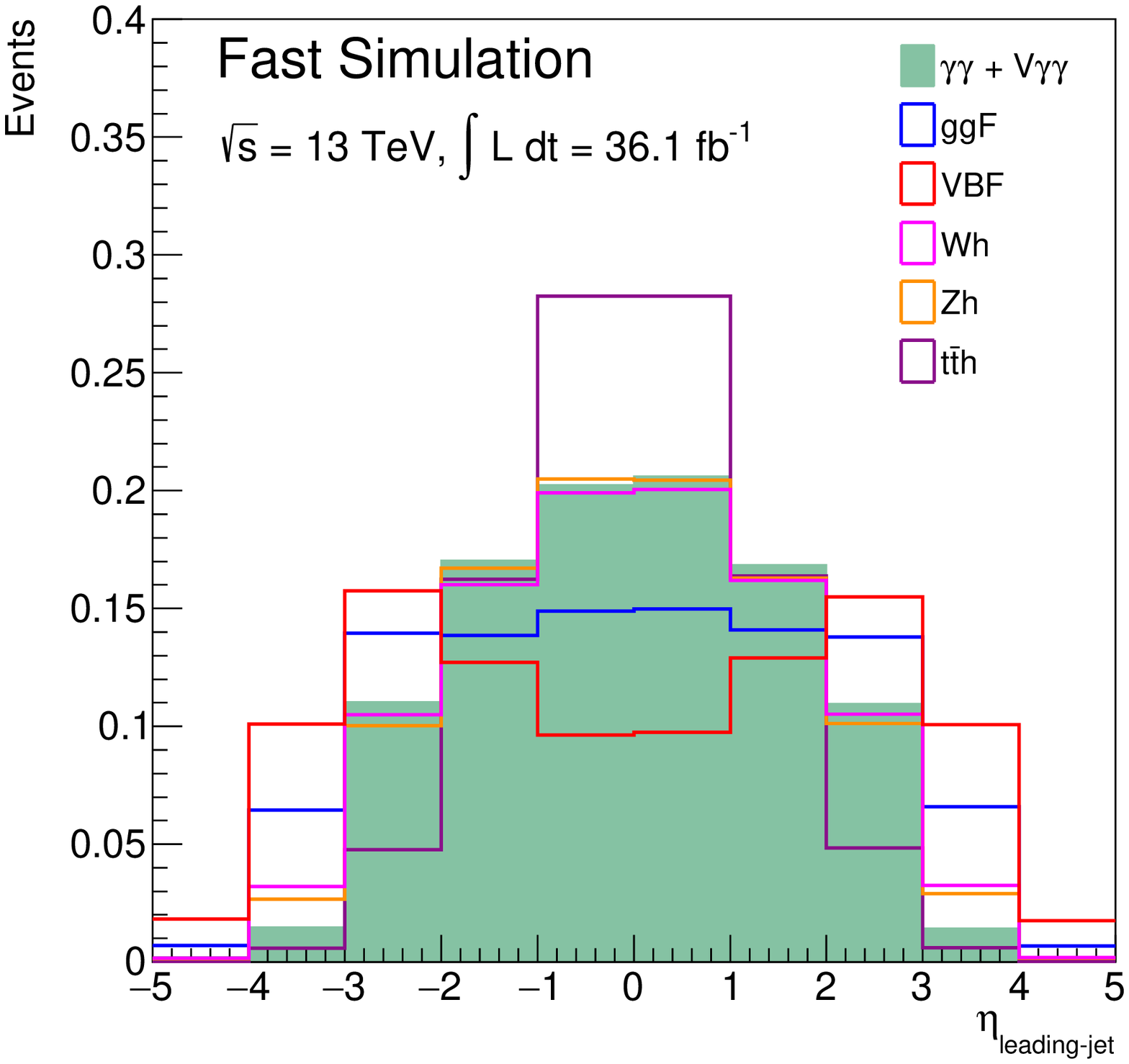}}
\subfigure[]{\includegraphics[width=0.4\linewidth]{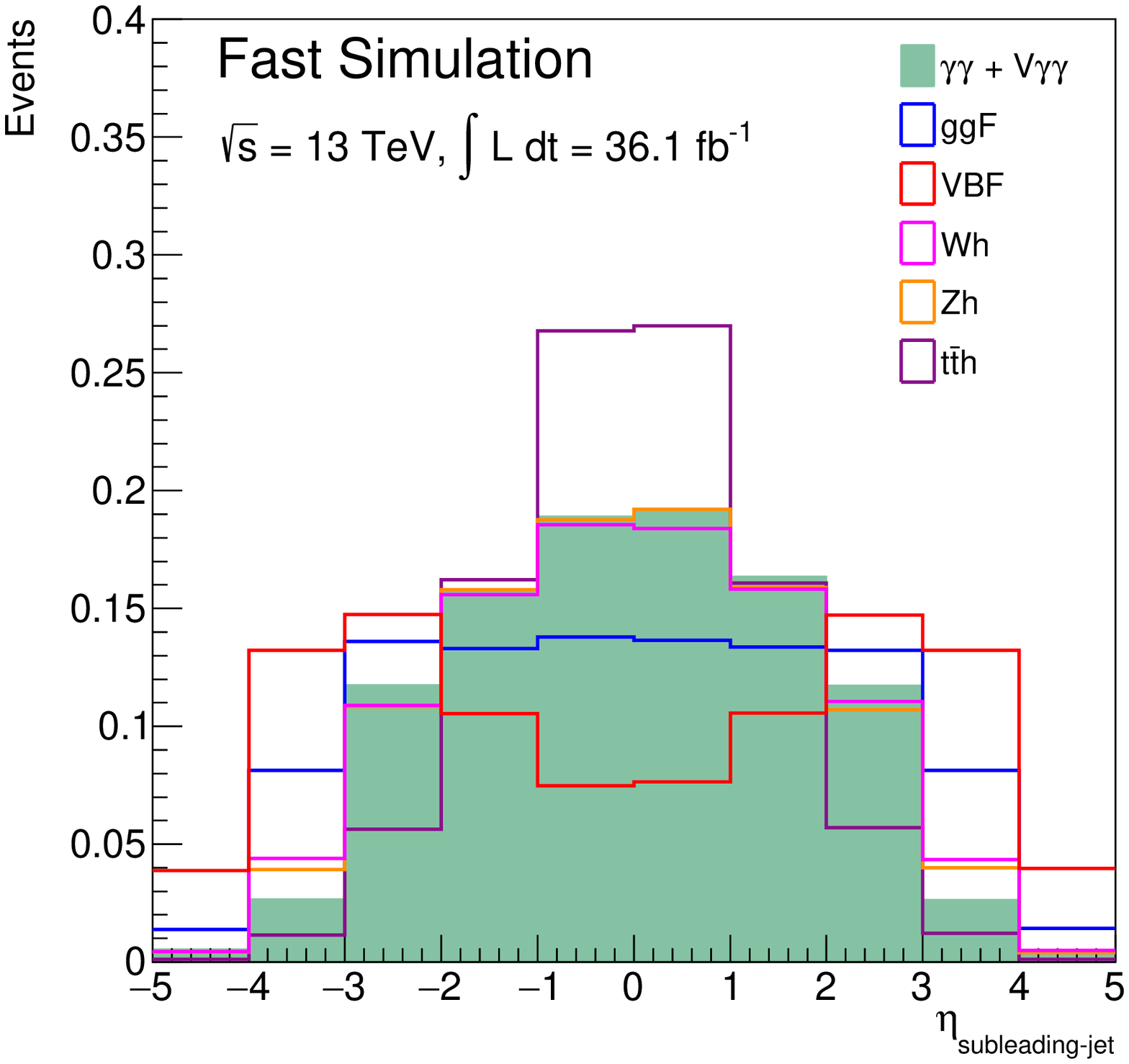}} \\
\subfigure[]{\includegraphics[width=0.4\linewidth]{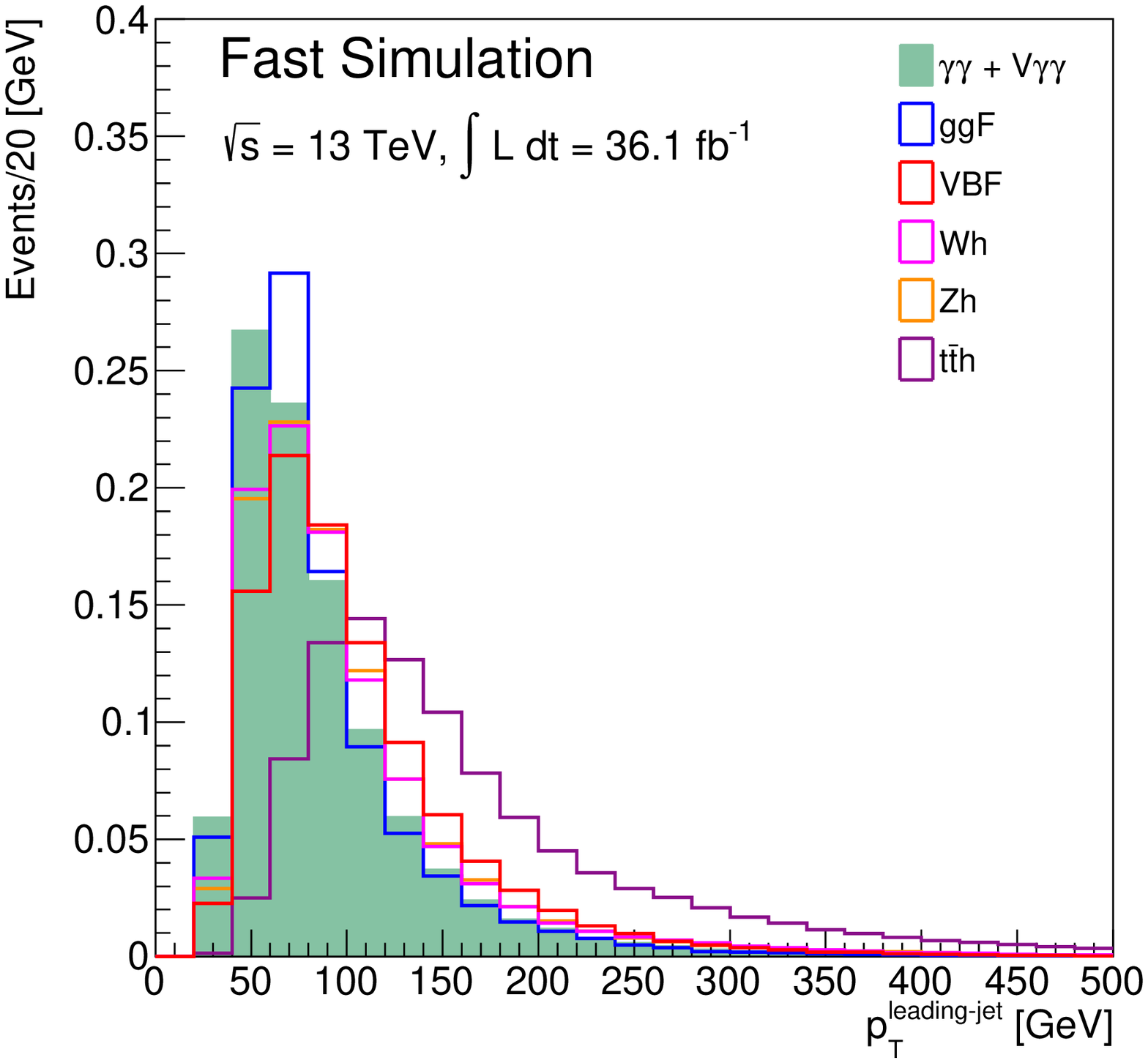}}
\subfigure[]{\includegraphics[width=0.4\linewidth]{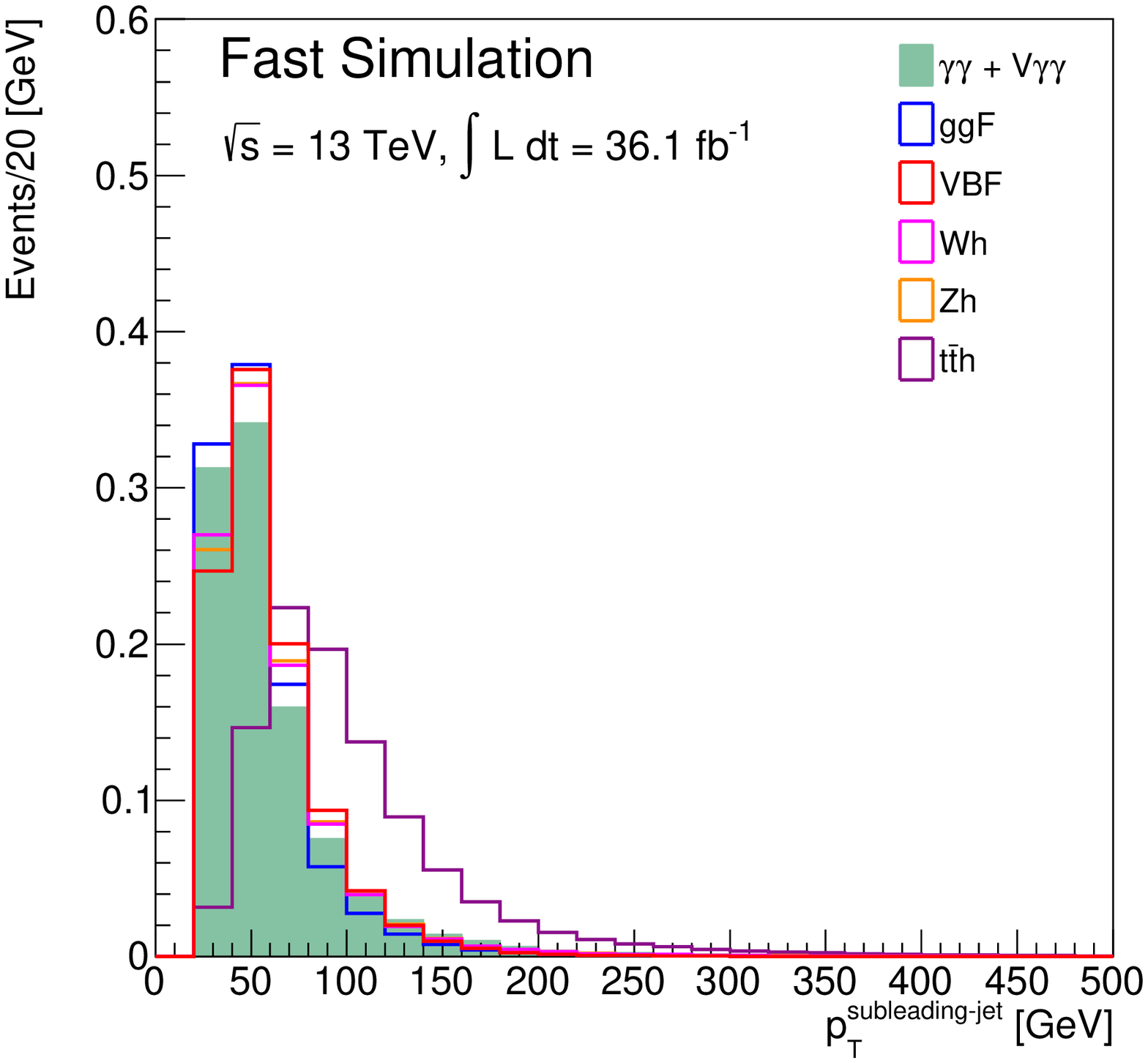}} \\
\subfigure[]{\includegraphics[width=0.4\linewidth]{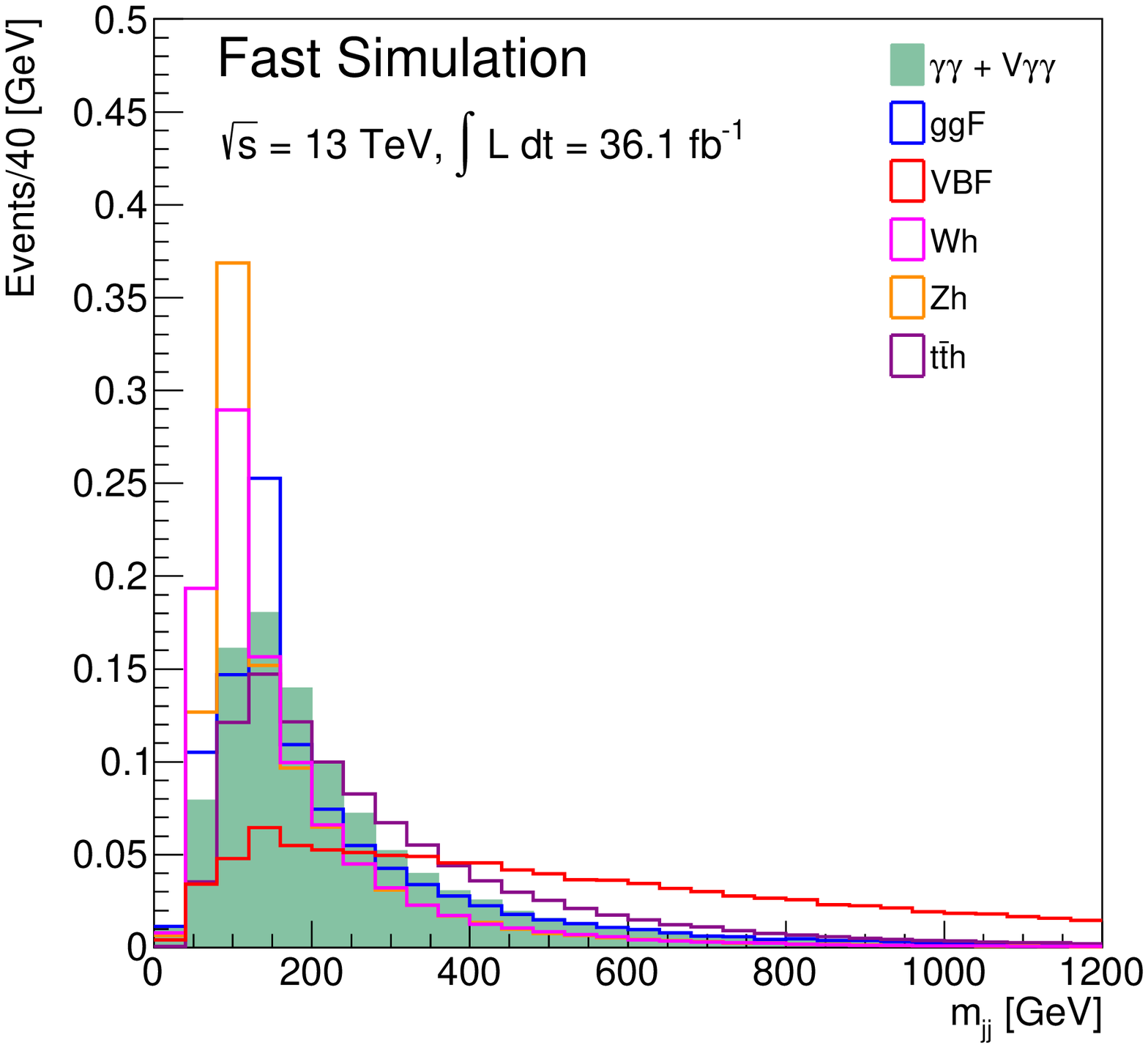}}
\subfigure[]{\includegraphics[width=0.4\linewidth]{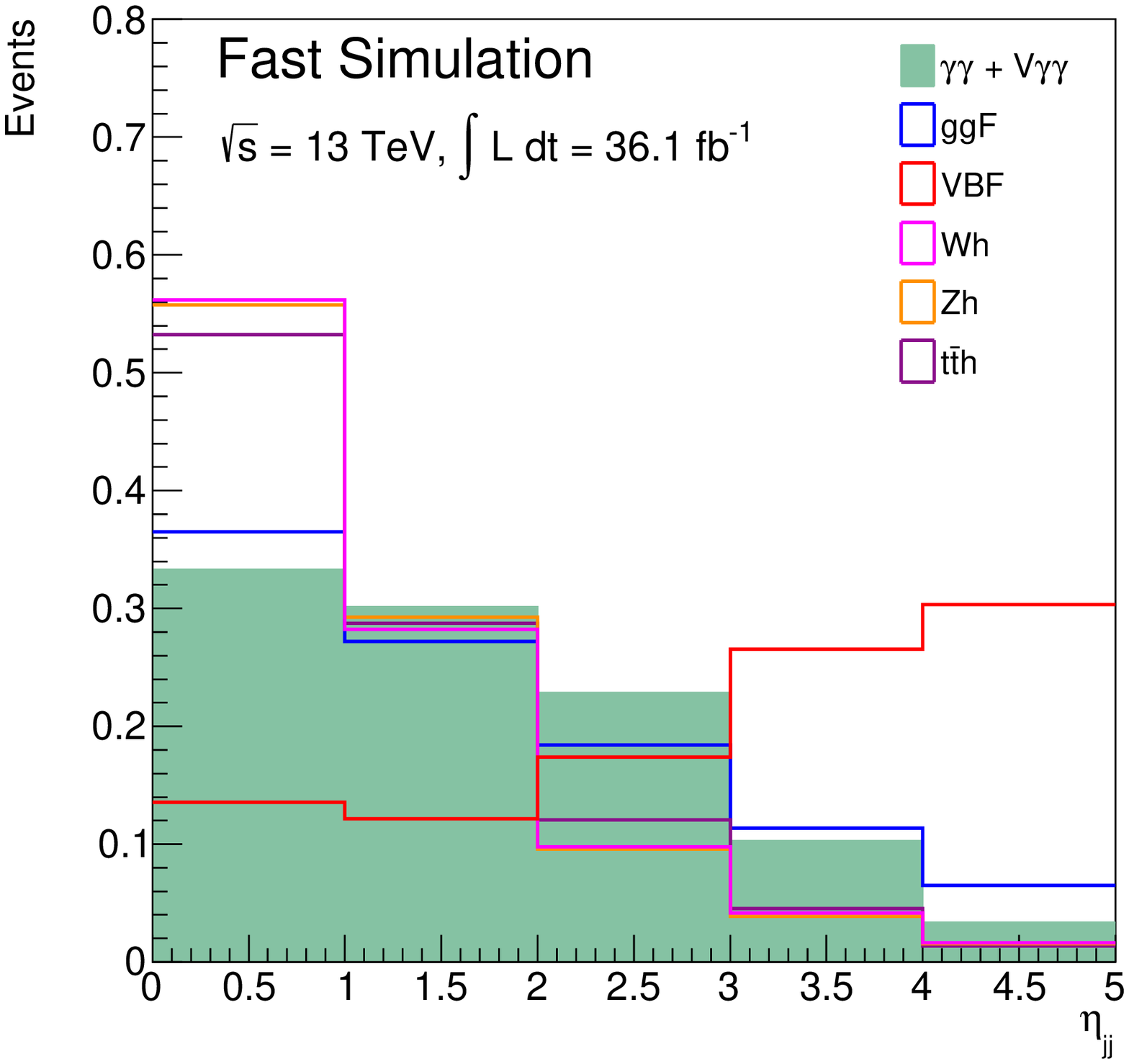}}
\caption{Signal and background distributions of the: (a) pseudorapidity of the leading jets, (b) pseudorapidity of the sub-leading jets, (c) transverse momentum of the leading jets, (d) transverse momentum of the sub-leading jets, (e) pseudorapidity separation between the leading and sub-leading jets of and (f) invariant mass of the leading and sub-leading system }\label{fig1:Input2}
\end{figure}

Processes with high transverse missing energy, such as $Wh$ and ${t\bar{t}h}$ are isolated using the reconstructed missing transverse energy, azimuthal angle separation between the leading jets and the missing transverse energy, the azimuthal angle separation between the di-photon system and the missing transverse energy and the azimuthal angle separation between the di-photon system and the leading jets, as displayed in Fig.~\ref{fig1:Input3}. Production mechanisms with at least two jets at the central region ($t\bar{t}h$) or the forward region (VBF, $Wh$ and $Zh$) of the detector are discriminated using the jet pseudorapidity, $\eta_{jet}$. The transverse momentum, $p_{jet}^{T}$, is used to identify processes with hard or soft jets. The di-jet mass, $m_{jj}$, and the pseudorapidity separation, $\Delta\eta_{jj}$, where used to characterise the VBF from the other processes. Figure~\ref{fig1:Input2} represents the distributions of pseudorapidity, transverse momentum of the leading and sub-leading jets, as well as the pseudorapidity separation between the leading and sub-leading jets and its invariant di-jet mass.

\FloatBarrier

\FloatBarrier
\newpage
\section{Effect of signal event yield on the event classification}

\begin{figure}[hbt!]
\centering
\subfigure[]{\includegraphics[width=0.42\linewidth]{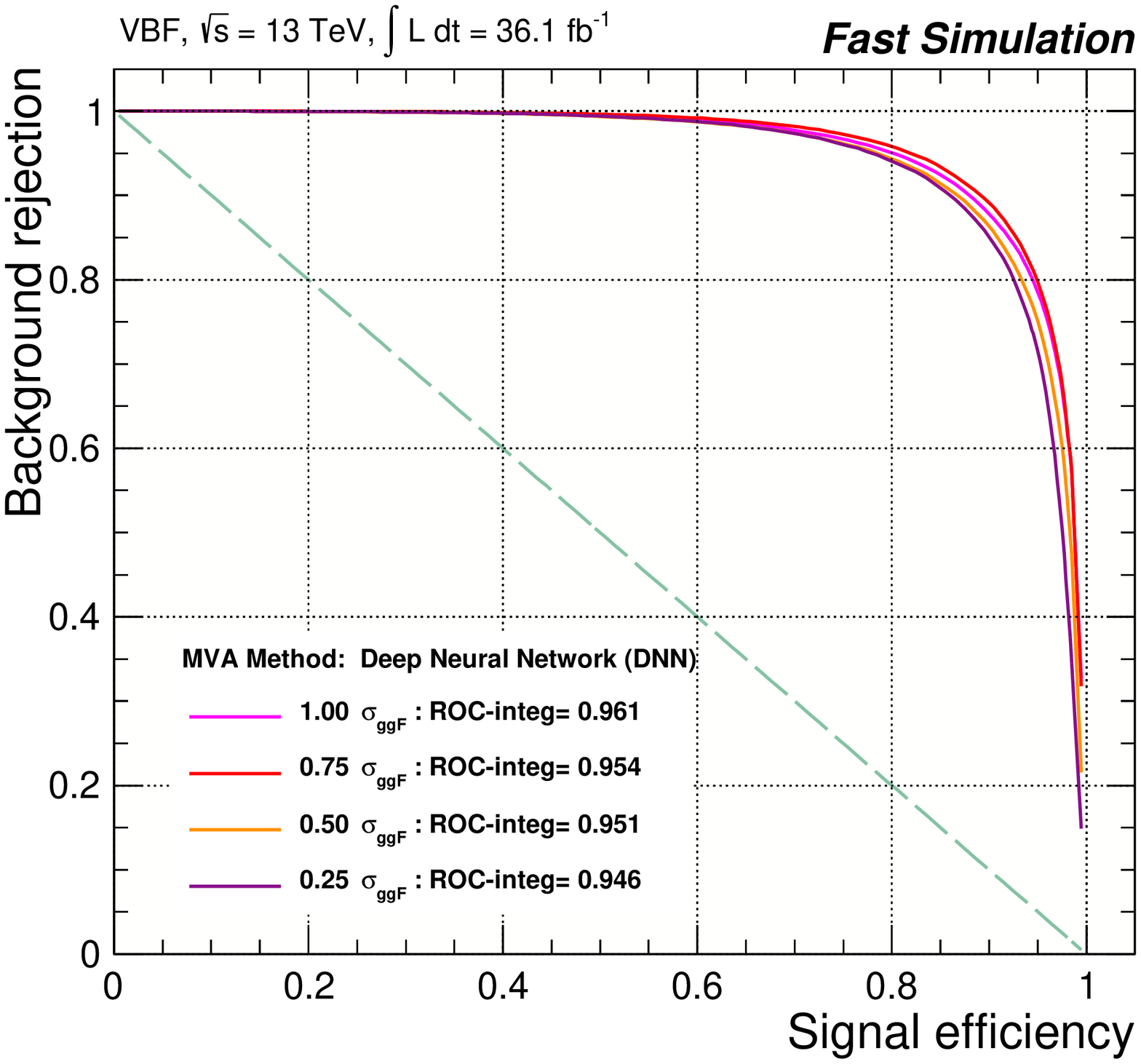}}
\subfigure[]{\includegraphics[width=0.42\linewidth]{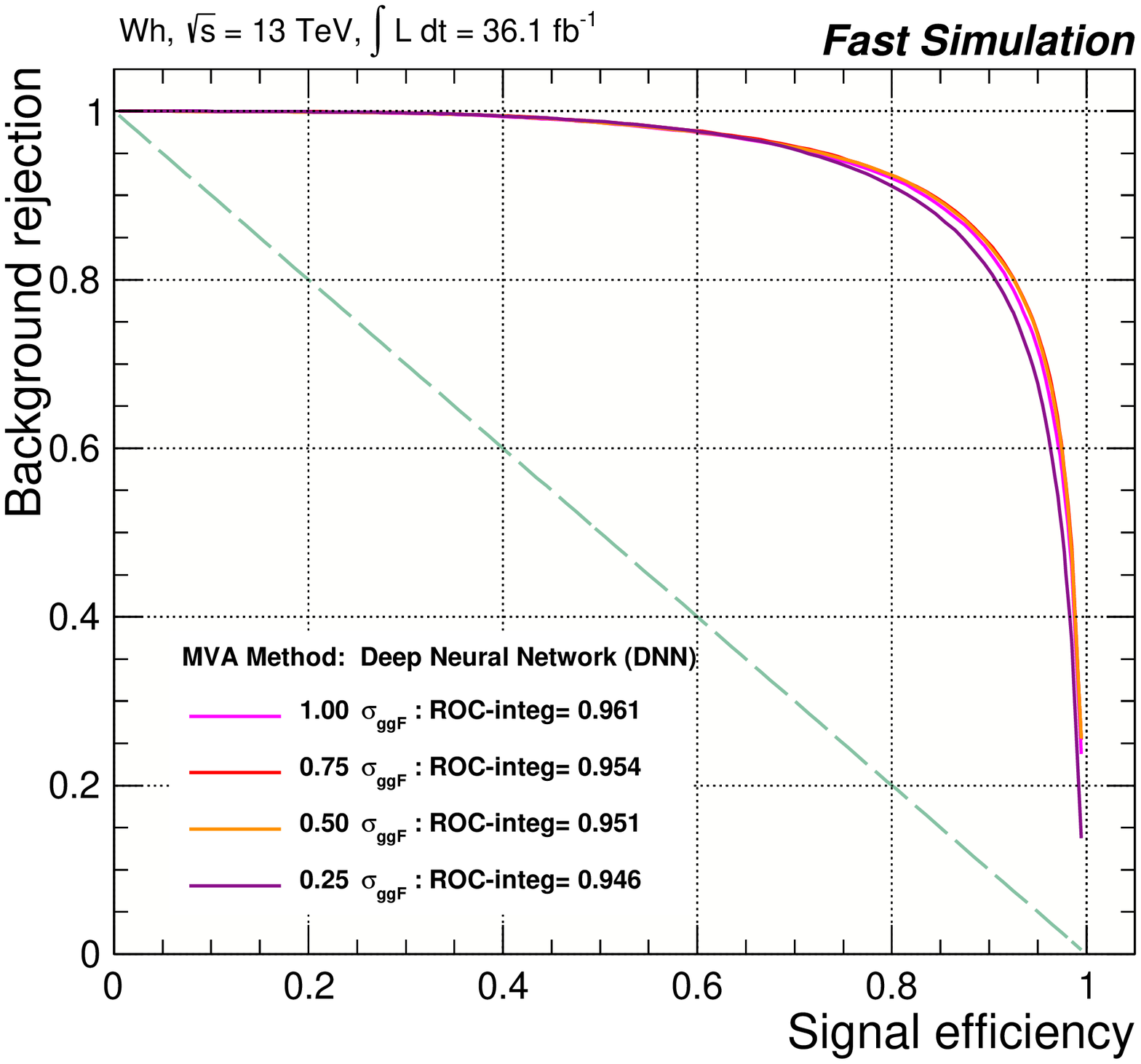}} \\
\subfigure[]{\includegraphics[width=0.42\linewidth]{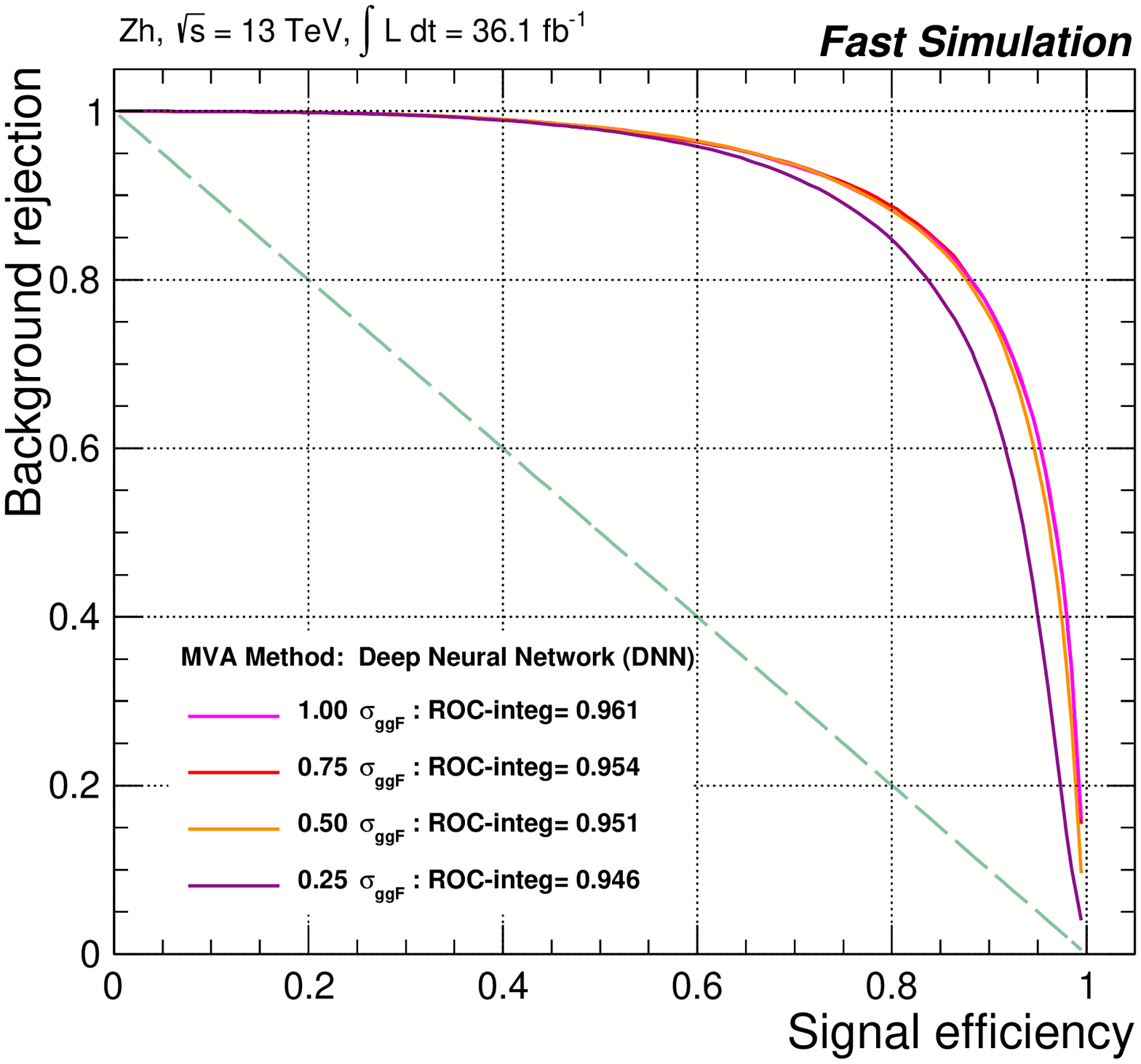}}
\subfigure[]{\includegraphics[width=0.42\linewidth]{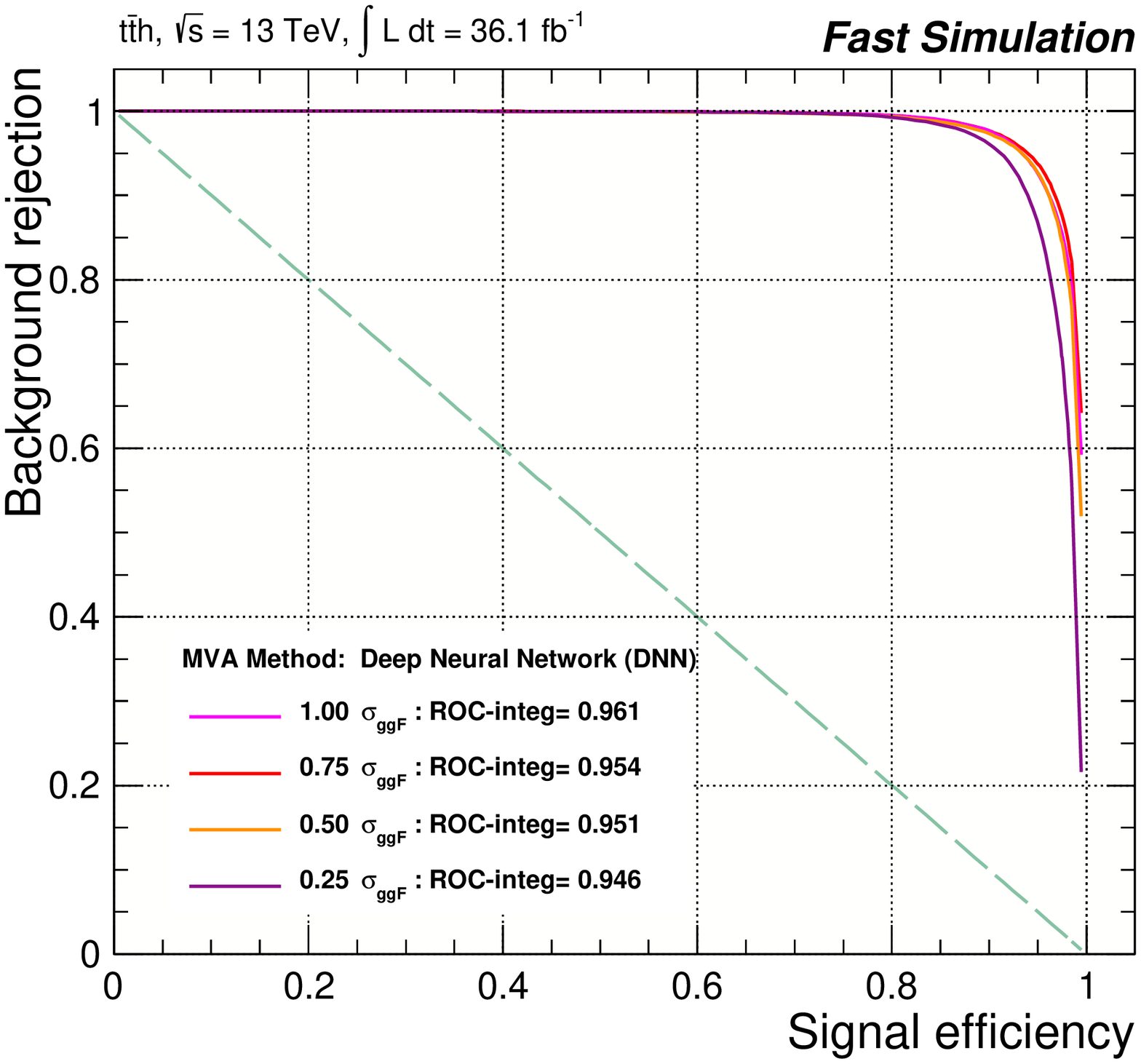}}
\caption{ ROC curves from supervised learning with unlabelled signals, (a) VBF, (b) $Wh$, (c) $Zh$ and (d) $t\bar{t}h$ are normalised to different fractions of the expected events from ggF production mechanism.}\label{fig1:StatROC}
\end{figure}

In order to assess the effect of the event yield that different production mechanisms have on the DNN classifier, unlabelled supervision is used. All the production mechanisms are grouped under the same label and have an expected cross-section normalised to 36.1~fb$^{-1}$. The expected cross-sections of VBF, $Wh$, $Zh$ and $t\bar{t}h$ are increased uniformly by the following fractions of the ggF process yield; 0.25, 0.5, 0.75 and 1. Figure~\ref{fig1:StatROC} shows the ROC curves of each of the production mechanisms, comparing the effects of increasing ggF cross-section on the signal efficiency. The results indicate that the effectiveness of an unlabelled classifier, depends on the relative contribution of each mechanism and the complexity of their topological features.
The success of the weak supervision classifier is similarly influenced by the contributions of production mechanisms and complexity of topological features.

\label{sec:Statist}
\FloatBarrier

\thispagestyle{plain}
\bibliographystyle{ws-ijmpa}
\bibliography{ws-ijmpa.bib}
\end{document}